\documentclass{aa} 
\usepackage{graphicx}
\usepackage{txfonts}
\usepackage{color}
\usepackage{ulem}
\usepackage{tablefootnote}
\usepackage{float} 
\usepackage{longtable}
\usepackage[labelfont=small,bf]{caption}
\usepackage{amsmath}
\usepackage[english]{babel}
\usepackage{comment}
\usepackage{multirow}
\usepackage{wasysym}
\usepackage{hyperref}
\usepackage{natbib}
\usepackage{lscape}
\usepackage{tabularx}
\usepackage{xtab,booktabs}

\bibpunct{(}{)}{;}{a}{}{,} % to follow the A&A style
\hypersetup{colorlinks=true,citecolor=blue}

\begin{document}
 \title{Multiplicity of stars with planets in the solar neighbourhood\thanks{Tables B.1, B.2, B.3, and B.4 are only available in electronic form at the CDS via anonymous ftp to cdsarc.u-strasbg.fr (130.79.128.5) or via \url{http://cdsweb.u-strasbg.fr/cgi-bin/qcat?J/A+A/}.}}

% \subtitle{}

 \titlerunning{Multiplicity of stars with planets in the solar neighbourhood}

 \authorrunning{J.~González-Payo et al.}
 
 \author{J.~González-Payo\inst{1,2}, J.\,A.~Caballero\inst{3}, J.~Gorgas\inst{1}, M.~Cortés-Contreras\inst{1}, M.-C.~Gálvez-Ortiz\inst{3}, C.~Cifuentes\inst{3}}

 \institute{Departamento de Física de la Tierra y Astrofísica \& IPARCOS-UCM (Instituto de Física de Partículas y del Cosmos de la UCM), Facultad de Ciencias Físicas, Universidad Complutense de Madrid, 28040 Madrid, Spain
  \email{fcojgonz@ucm.es}
 \and
 UNIE Universidad, Departamento de Ciencia y Tecnología, c/ Arapiles 14, 28015 Madrid, Spain
 \and
 Centro de Astrobiología (CSIC-INTA), European Space Astronomy Centre, Camino Bajo del Castillo s/n, 28692 Villanueva de la Cañada, Madrid, Spain
  }

 \date{Received 20 March 2024 / Accepted 29 July 2024}
 
 \abstract
 {}
 {We intended to quantify the impact of stellar multiplicity on the presence and properties of exoplanets.
 }
 {We investigated all exoplanet host stars at less than 100\,pc using the latest astrometric data from \textit{Gaia} DR3 and advanced statistical methodologies.
 We complemented our search for common proper motion and parallax companions with data from the Washington Double Star catalogue and the literature.
 After excluding a number of systems based on radial velocity data, and membership in clusters and open associations, or with resolved ultracool companions, we kept 215 exoplanet host stars in 212 multiple-star systems.
 }
 {We found 17 new companions in the systems of 15 known exoplanet host stars, measured precise angular and projected physical separations and position angles for 236 % 261-25 
 pairs of stars, compiled key parameters for 276 planets in multiple systems, and established a comparison sample comprising 687 single stars with exoplanets.
 With all of this, we statistically analysed a series of hypothesis regarding planets in multiple stellar systems.
 Although they are only statistically significant at a $2\sigma$ level, our analysis pointed to several interesting results on the comparison in the mean number of planets in multiple versus single stellar systems and the tendency of high mass planets to be located in closer orbits in multiple systems.
 We confirm that planets in multiple systems tend to have orbits with larger eccentricities than those in single systems. In particular, we found a significant ($> 4\sigma$) preference for planets to exhibit high orbital eccentricities at small ratios between star-star projected physical separations and star-planet semi-major axes.
 }
 {}

  % stars: kinematics and dynamics

 \maketitle

\section{Introduction}
\label{sec:introduction}

After over 30 years since the discovery of the first exoplanets \citep{campbell88,wolszczan92,mayor95}, almost 6000 exoplanet candidates have been reported. 
Except for a few ``rogue'' planets \citep[and references therein]{caballero18} and bodies of unknown nature found around compact objects such as white dwarfs and neutron stars \citep{bailes11,vanderburg20}, the majority of these exoplanet candidates orbit around main-sequence and giant stars.
Planet occurrence rates determined with different techniques indicate that, on average, there are slightly more than one planet per star \citep{cassan12,dressing15,baron19,savel20,sabotta21,ribas23}.
Additionally, a significant fraction of the stars in the Galaxy are part of stellar multiple systems, including double, triple, or higher-order systems \citep[][and many others]{abt76,duchene13,tokovinin08}.
As a consequence, many of the discovered exoplanet candidates are also part of stellar multiple systems.
 
\begin{table*}
 \centering
 \caption[]{Observational works on multiplicity of stellar systems with planets.}
 \scalebox{0.85}[0.85]{
 \begin{tabular}{llcc}
 \hline \hline
 \noalign{\smallskip}
 Title & Reference & $N^{(a)}$ & Methodology$^{(b)}$ \\
 \noalign{\smallskip}
 \hline
 \noalign{\smallskip} 
 Multiplicity of stars with planets in the solar neighbourhood & This work & 215  &  Misc. \\
 An early catalog of planet-hosting multiple-star systems of order three...  & \citet{cuntz22} & 40 & Misc. \\
 Speckle observations of TESS exoplanet host stars. II. Stellar...  & \citet{lester21} & 102 &  SI \\   
 The census of exoplanets in visual binaries: population trends from a...  & \citet{fontanive21} & 218 & Misc. \\ 
 How many suns are in the sky? A SPHERE multiplicity... I. Four new ...  & \citet{ginski21} & 4 & AO \\ 
 Frequency of planets in binaries & \citet{bonavita20} & 313 & RV \\
 Understanding the impacts of stellar companions on planet formation... & \citet{hirsch21} & 109 & AO, RV \\
 Search for stellar companions of exoplanet host stars by exploring... & \citet{mugrauer19} & 204 & WFI \\
 The SEEDS high-contrast imaging survey of exoplanets around young... & \citet{uyama17} & 68 & AO \\
 A lucky imaging multiplicity study of exoplanet host stars - II  & \citet{ginski16} & 60 & LI \\
 High-contrast imaging search for stellar and substellar companions of... & \citet{mugrauer15} & 33 & AO \\
 A Lucky Imaging search for stellar companions to transiting planet host... & \citet{wollert15} & 49 & LI \\
 Binary frequency of planet-host stars at wide separations. A new brown...  & \citet{lodieu14} & 37 & WFI \\
 The multiplicity status of three exoplanet host stars & \citet{ginski13} & 3 & LI \\
 Stellar companions to exoplanet host stars: Lucky imaging of transiting... & \citet{bergfors13} & 21 & LI  \\
 Extrasolar planets in stellar multiple systems  & \citet{roell12} & 57 &  AO \\
 Know the star, know the planet. I. Adaptive optics of exoplanet host...  & \citet{roberts11} & 62 & AO  \\
 Know the star, know the planet. II. Speckle interferometry of exoplanet...  & \citet{mason11} & 118 &  SI \\
 Binarity of transit host stars. Implications for planetary parameters & \citet{daemgen09} & 14 & LI \\
 The frequency of planets in multiple systems  & \citet{bonavita07} & 202 & Misc. \\
 The multiplicity of exoplanet host stars. New low-mass stellar companions & \citet{mugrauer09} & 43 & WFI \\
 The multiplicity of planet host stars - new low-mass companions to planet...  & \citet{mugrauer07} & 3 &  Misc. \\
 The impact of stellar duplicity on planet occurrence and properties. I. ...  & \citet{eggenberger07} & 57 & AO \\   
 A search for wide visual companions of exoplanet host stars: The Calar ... & \citet{mugrauer06} & 44 & WFI \\
 Two suns in the sky: Stellar multiplicity in exoplanet systems & \citet{raghavan06} & 36 & Misc.  \\
 Detection and properties of extrasolar planets in double and multiple... & \citet{eggenberger04} & 15 & AO \\
 Planets in multiple-star systems: properties and detections & \citet{udry04} & 15 & Misc. \\
Stellar companions to stars with planets & \citet{patience02b} & 11 & Misc. \\
 \noalign{\smallskip}
 \hline
 \end{tabular}
 }
 \label{tab:studies} 
   \tablefoot{
    \tablefoottext{a}{$N$: number of investigated multiple stellar systems with exoplanets.}
    \tablefoottext{b}{AO: Adaptive optics; LI: Lucky imaging; RV: Radial velocity; SI: Speckle imaging; WFI: Wide field imaging; Misc.: Miscellanea.}} 
\end{table*}

Stellar multiplicity has been studied for centuries \citep{mayer1778,herschel1802,duquennoy91,jao09a,duchene13}.
In our immediate vicinity, that is, in the 10\,pc-radius volume centred on our Sun, \citet{reyle21} measured an overall multiplicity fraction (MF) of $27.4 \pm 2.3$\,\%.
Actually, the MF decreases from the earliest to the latest spectral types:
it varies from virtually 100\% for O stars (except for a few runaway stars, they are located in dense star-forming regions), through more than 70\% for B and A stars \citep{kouwenhoven07b,mason09,sana13,caballero14,maizapellaniz19},
44--67\% for solar-type stars \citep{duquennoy91,raghavan10,duchene13}, 26--40\% for M dwarfs \citep{fischer92,reid97c,delfosse04,janson12,ward15,cortes17b,winters19}, to below 20\% for L, T, and Y ultracool dwarfs \citep{burgasser03c,burgasser05,burgasser07b,fontanive18}.
Since most of the reported exoplanet candidates orbit FGKM-type stars, several thousands of planetary systems would also be expected among multiple stellar systems.
This fact is actually modulated by an observational bias by which many exoplanet surveys tend to discard close binaries in their input catalogues (e.g. \citealt{raghavan06,winn15,sebastian21,ribas23}).
Currently, there are a few hundreds multiple stellar systems with known exoplanet candidates \citep[][and many others]{eggenberger04b,konacki05,furlan17,martin18,bonavita20}. 

The consequence for a star being in a multiple system is not only relevant for the formation and evolution of the star itself, but also in terms of the formation and evolution of planets, especially if the stellar companions are close to each other.
Previous theoretical works had predicted that nearby companion stars can significantly disrupt circumstellar discs and hinder the process of planetary formation 
(\citealt{lissauer87,gladman93,jensen96,artymowicz96,pichardo05,quintana06,hamers21}; see also \citealt{thebault15} and the series of papers initiated by \citealt{wang14a}),
while other theoretical works have focused on the long-term stability of planets in binary systems. 
For example, \citet{holman99} were among the first to study in great detail such stability.
A number of similar works on the long-term stability of S- (around one star) and P-type (around the two stars) systems have been published afterwards by, e.g. \citet{pilatlohinger03}, \citet{musielak05}, \citet{mudryk06}, and \citet{doolin11}.
There has also been theoretical work on long-term stability of triple systems with planets \citep[e.g.][]{busetti18}.
Most of these publications, however, have focused on the eccentricity of stellar orbits, rather than on the eccentricity of planet orbits.
The relation between stellar multiplicity and planet eccentricity was first investigated by \citet{mazeh97}, who analysed the stability of the 16~Cyg system.
It contains three stars (A: G1.5\,V; B: G3\,V; C: mid-M) and a 1.8\,M$_{\rm Jup}$-mass planet in an eccentric orbit ($e \approx$ 0.68) around the secondary \citep{cochran97,turner01,rosenthal21}.

Likewise, there has been a number of observational work on multiplicity of stellar systems with planets.
Table~\ref{tab:studies} enumerates many relevant observational publications on this topic.
The authors have used a diversity of target stars with planets, methodologies, and even maximum distances (e.g. \citealt{hirsch21}: 25\,pc; \citealt{eggenberger07}: 50\,pc; \citealt{fontanive21}: 200\,pc; \citealt{mugrauer19,lester21}: 500\,pc). 
A wealth of results have been proposed after these observations.
For example, \citet{eggenberger04} and \citet{udry04} found that massive planets ($M \sin{i} >$ 2\,M$_{\rm Jup}$) with moderately short periods ($P \le$ 40--100\,d) in binaries tend to have low eccentricities, while \citet{moutou17}, with a more powerful facility (SPHERE at the Very Large Telescope), also concluded that the majority of high-eccentricity planets are ``not embedded'' in multiple stellar systems.
These results are rather inconsistent with those of \citet{raghavan06}, who showed that planets in systems with confirmed stellar companions generally have instead higher eccentricities.
They reasoned that companion stars would have a greater gravitational influence on the planets' orbits and could shorten the periods of Kozai cycles \citep{innanen97,wu03,tamuz08,naoz16}.
We finish this introduction with a sentence written by \citet{eggenberger04} exactly two decades ago and that is still valid:
``{The studies [enumerated above] emphasise the importance of searching for planets in multiple star systems, even though it is more challenging to carry out than the search for planets around individual stars}''.

In our work, we revisit the topic of multiplicity of stars with planets at less than 100\,pc.
After this introduction (Sect.~\ref{sec:introduction}), we describe our target sample in Sect.~\ref{sec:sample}. 
Next, we detail in Sect.~\ref{sec:analysis} our analysis, which consists in an individualised search for common proper-motion and parallax companions to exoplanet host stars with \textit{Gaia} DR3 complemented with data from the Washington Double Star catalogue and the literature.
The results are presented in Sect.~\ref{sec:results_discussion}, where we report on new stellar systems with planets, the relationships between star-star separations, star-planet semi-major axes, and planet eccentricities, and number and masses of planets in single and multiple systems, and conclude on the properties of multiple star systems with planets.
Finally, Sects.~\ref{sec:discussion} and~\ref{sec:summary} discuss and summarise our results.

\section{Sample}
\label{sec:sample}

\begin{figure}
 \centering
 \includegraphics[width=1\linewidth, angle=0]{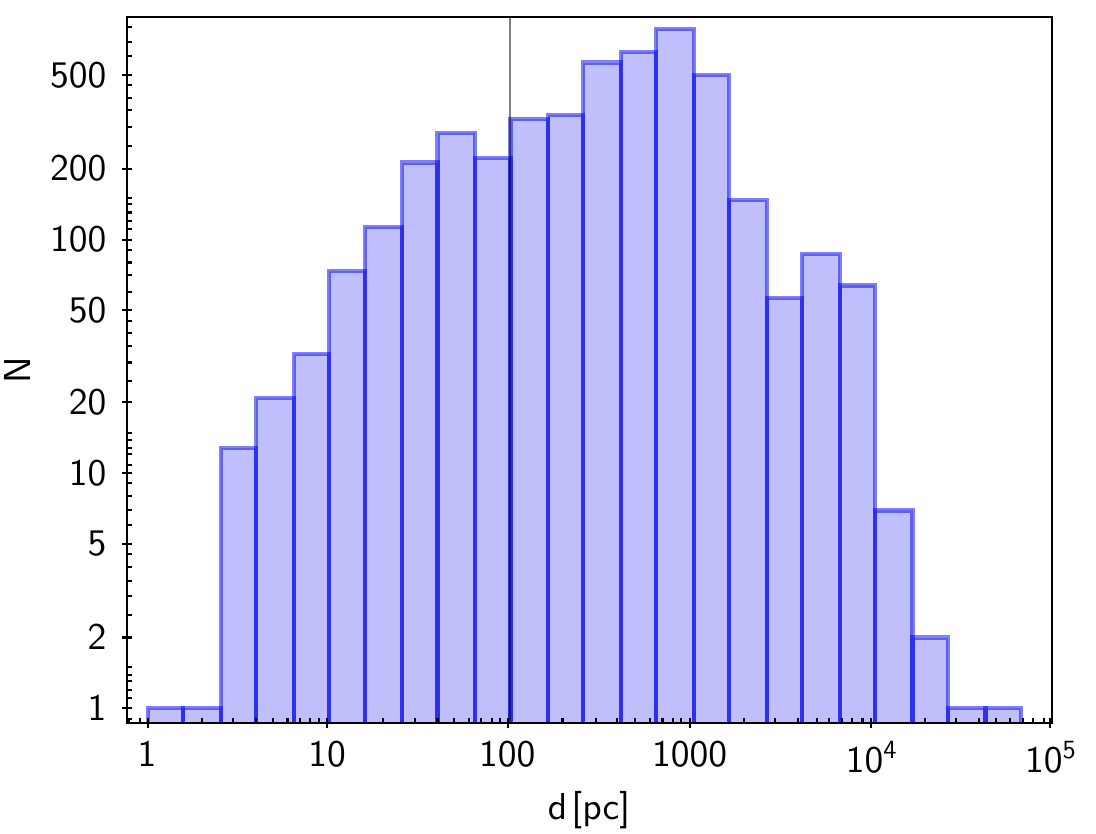}
 \includegraphics[width=1\linewidth, angle=0]{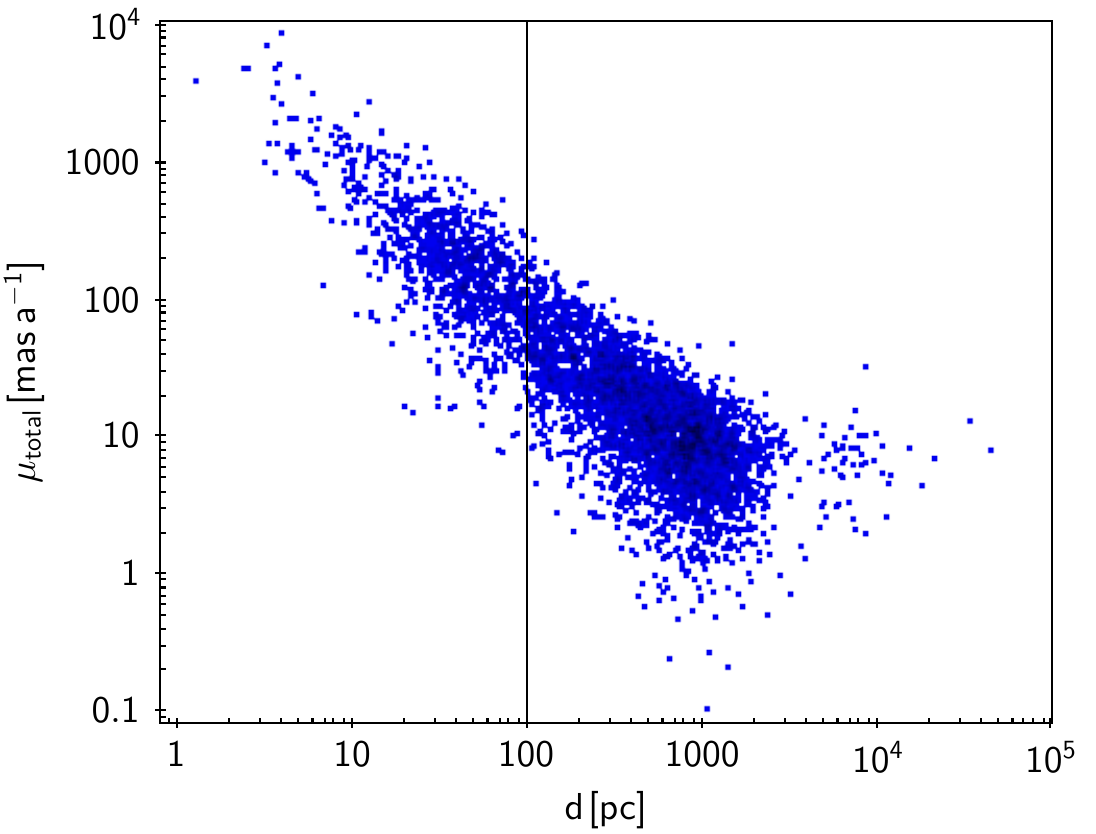}
 \caption{\textit{Top panel}: Histogram of distances of the 4583 exoplanet host stars of the initial sample.
 \textit{Bottom panel}: Total proper motion as a function of distance for the same stars.
 The vertical line at $d$ = 100\,pc in both panels indicate the search limit.
}
 \label{fig:distances}
\end{figure}

Our sample was built on the basis of the two most widely used exoplanet databases. We took all the host stars tabulated by either the Extrasolar Planets Encyclopaedia\footnote{Now Encyclopaedia of Exoplanetary Systems, \url{http://exoplanet.eu/}} \citep{schneider11} or the NASA Exoplanet Archive\footnote{\url{https://exoplanetarchive.ipac.caltech.edu/}} \citep{akeson13}. At the moment of downloading (3 January 2024), the first database contained 5576 planets in 4114 planetary systems, and the second one had 5566 planets in 4145 systems. 
Firstly, we discarded all of the duplicated host stars in both databases. Sometimes, the same star was tabulated with different names in each database, and not always had the same exact coordinates. 
We finally obtained a set of 4612 non-duplicated host stars, all of which have an entry in the Simbad astronomical database \citep{wenger00}.

Secondly, we looked for the \textit{Gaia} DR3 \citep{gaiacollaboration22} counterpart of every host star. 
For some cases that required a visual inspection, we used the Aladin Sky Atlas \citep{bonnarel00}. 
Of the 4612 host stars, 339 do not have a \textit{Gaia} DR3 entry. 
Of them, 8 are just too bright for {\it Gaia}\footnote{The 8 very bright exoplanet host stars are: $\alpha$~PsA (Fomalhaut), $\alpha$~Ari (Hamal), $\alpha$~Tau (Aldebaran), $\beta$~And (Mirach), $\beta$~Gem (Pollux), $\beta$~UMi (Kochab), $\gamma$~Lib (Zubenelhakrabi), and $\gamma^{01}$~Leo.}, while the other 331 stars without a \textit{Gaia} DR3 entry are distant microlensing objects from the OGLE\footnote{Optical Gravitational Lensing Experiment, \url{https://ogle.astrouw.edu.pl/}} \citep{udalski92}, KMT\footnote{Korea Microlensing Telescope Network, \url{https://kmtnet.kasi.re.kr/kmtnet/}} \citep{henderson14}, and MOA\footnote{Microlensing Observations in Astrophysics, \url{http://www2.phys.canterbury.ac.nz/moa/}} \citep{alcock95,alcock97} surveys, and some pulsars. 

Of the 4273 stars with an entry in \textit{Gaia} DR3, 4232 have a positive parallax value. 
For the remaining 41 stars (with negative parallax or no parallax at all) plus the former 339 with no counterpart in \textit{Gaia} DR3, we looked for their parallaxes or spectro-photometric distances in the \textit{Hipparcos} \citep{perryman97}, \textit{Gaia} DR2 \citep{gaiacollaboration18}, \textit{Gaia} EDR3 \citep[Third Early Data Release --][]{gaiacollaboration21},
and UCAC4 \citep[Fourth United States Naval Observatory CCD Astrograph Catalogues --][]{zacharias13}
catalogues and in the literature (e.g. \citealt{shkolnik12,gagne15,leggett17,finch18,winters19}). % Chauvin et al. 2011
We were able to find a parallax or a distance for 351 of the mentioned 380 stars. 
To sum up, a total of 4583 exoplanet host stars in our sample have a tabulated distance, and only 29 (0.63\% of the 4612 non-duplicated host stars) have not.
The distribution of distances of the 4583 stars is shown in the top panel of Fig.~\ref{fig:distances}.

Next, we restricted the analysis to stars with distances less than 100\,pc, which is the limit of the solar neighbourhood \citep{gaiacollaboration21b}. 
There are many practical reasons behind this search limit selection, such as manageability of the final sample for detailed analysis, increase of both the incompleteness of the exoplanet searches and of the astrometric uncertainty for the search for common proper motion companions at longer distances, as illustrated by the bottom panel of Fig.~\ref{fig:distances}, and reliability of the parameters of the planetary systems, which have been mostly detected with the radial velocity and transit methods.
Finally, 998 non-duplicated exoplanet host stars are located at less than 100\,pc.
They are listed in Table~\ref{tab:sample}.

\section{Analysis}
\label{sec:analysis}

\begin{figure}
 \centering
 \includegraphics[width=0.96\linewidth, angle=0]{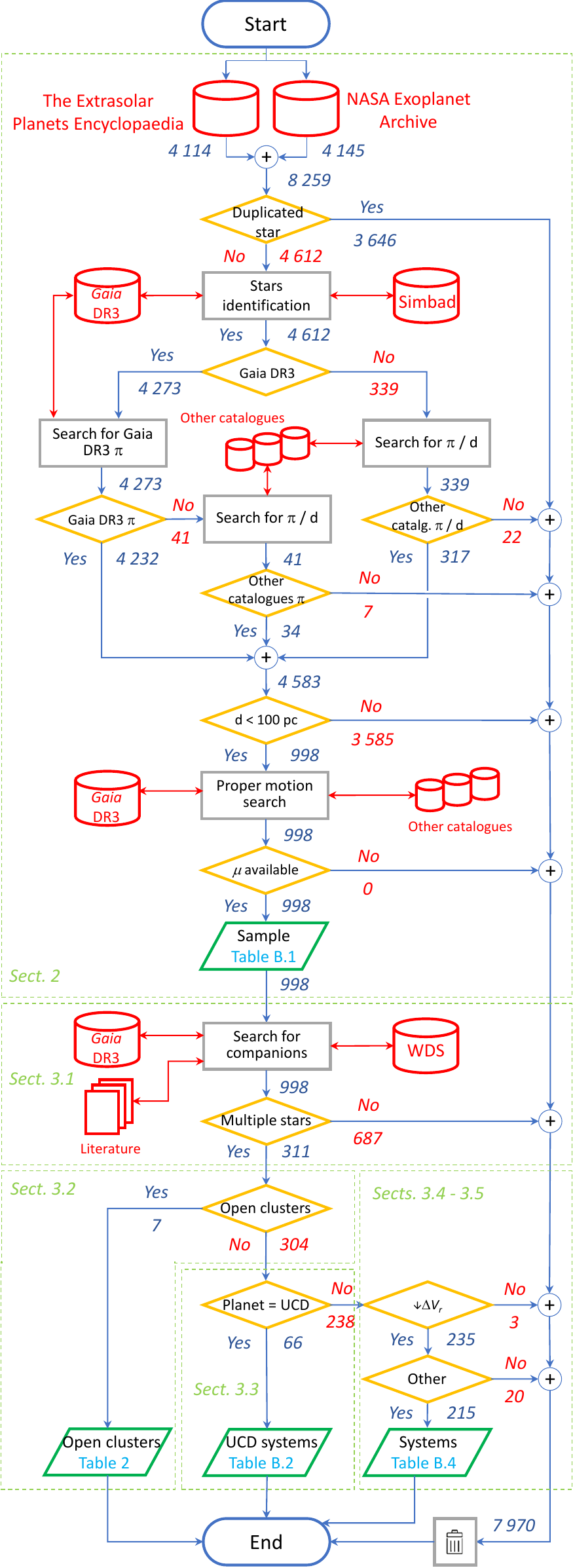}
 \caption{Flowchart describing the sample preparation and following analysis.}
 \label{fig:flowchart}
\end{figure}

The flowchart in Fig.~\ref{fig:flowchart} summarises the sample preparation and the following analysis.
In the diagram, the stadia indicate the beginning and ending, cylinders the databases, small circles the connectors, diamonds the decisions, rectangles the processes, stacks of rectangles the literature documents, rhomboids the inputting and outputting data, and flow lines (arrowheads) the processes' order of operation.
Every flowline is labelled with the corresponding number of host stars.

\subsection{Search for stellar companions}
\label{sec:search_for_stellar_companions}

The first step of the analysis was to look for companions of common \textit{Gaia} DR3 proper motion and parallax to our 998 non-duplicated exoplanet host stars at less than 100\,pc. 
We used the same methodology as \cite{gonzalezpayo21} (their Sect.~3).
In particular, we used \texttt{Topcat} \citep[Tool for OPerations on Catalogues And Tables;][]{taylor05} with a customised code in ADQL \citep[Astronomic Data Query Language;][]{yasuda04} to search for companions at projected physical separations, $s = \rho \cdot d$, of up to 1\,pc that satisfy the following criteria to distinguish between physical (bound) and optical (unbound) systems: 

\begin{equation}
\mu_\mathrm{ratio}=\sqrt{\frac{(\mu_{\alpha} \cos{\delta_1}-\mu_{\alpha} \cos{\delta_2})^2+(\mu_{\delta 1}-\mu_{\delta 2})^2}{(\mu_{\alpha} \cos{\delta_1})^2+(\mu_{\delta 1})^2}}<0.15,
\label{eqn:crit1}
\end{equation}

\begin{equation}
\Delta \text{PA}=\lvert  \text{PA}_1- \text{PA}_2 \rvert<15\,\mathrm{deg},
\label{eqn:crit2}
\end{equation}

\noindent and

\begin{equation}
\Biggl\lvert \frac{\pi_1^{-1}-\pi_2^{-1}}{\pi_1^{-1}} \Biggr\rvert < 0.15,
\label{eqn:crit3}
\end{equation}

\noindent where PA$_i$ is the angle between the proper motion vectors, with $i=1$ for the primary star and $i=2$ for the companion.
The inverse of the parallax, $\pi_i^{-1}$, is the distance, which for $d <$ 100\,pc in general does not need any further correction (e.g. \citealt{bailerjones18a,luri18a}). 
The motivation of the 0.15 and 15\,deg values by \cite{gonzalezpayo21} was justified by the dissimilarity of proper motions and parallaxes of bona fide physically bound stars of different \textit{Gaia} colours in close resolved systems with astrometric solutions (i.e., we neither applied a colour-term parallax correction nor subtracted relative motions in systems with orbital periods of a few years).
We found 230 pairs of stars that satisfy these criteria. 
While resolved binary systems are made of one pair of stars, resolved multiple systems are made of two (triple) or three (quadruple) pairs. 
See \citet{gonzalezpayo21} for further details on the search methodology.

Next, we complemented our \textit{Gaia} DR3 search for companions with a cross match with data in the Washington Double Star catalogue \citep[WDS;][]{mason01}.
Currently, they tabulate angular separations and position angles for the first and last epochs of observation of about 156\,000 multiple systems.
While they also tabulate other parameters (e.g. equatorial coordinates, magnitudes, notes on systems), additional information can be obtained from the WDS team upon request.

There are 687 WDS pairs in 341 systems containing at least one of the 998 input stars or the 230 \textit{Gaia} DR3 companions from our previous search.
Since there are ultra-wide multiple systems with extremely large projected separations \citep{caballero09,shaya11,gonzalezpayo23}, we kept WDS systems with $s >$ 1\,pc if they satisfy the three astrometric criteria (Eqs.~\ref{eqn:crit1}, \ref{eqn:crit2}, and \ref{eqn:crit3}).
The remaining stars, either host stars or \textit{Gaia} companions, are not part of any WDS system.

Of the 687 WDS pairs, we rejected 417 because their stars have very different \textit{Gaia} DR3 distances and proper motion moduli and direction (i.e. do not satisfy at least one of Eqs. \ref{eqn:crit1}, \ref{eqn:crit2}, and~\ref{eqn:crit3}).
Generally, the discarded companions are further in the background and have lower proper motions than the planet host stars.
Many of them have the WDS `U' flag: ``Proper motion or other technique indicates that this pair is non-physical''.
The other 270 WDS pairs passed to the next step of our analysis.

\begin{figure}
 \centering
 \includegraphics[width=1\linewidth, angle=0]{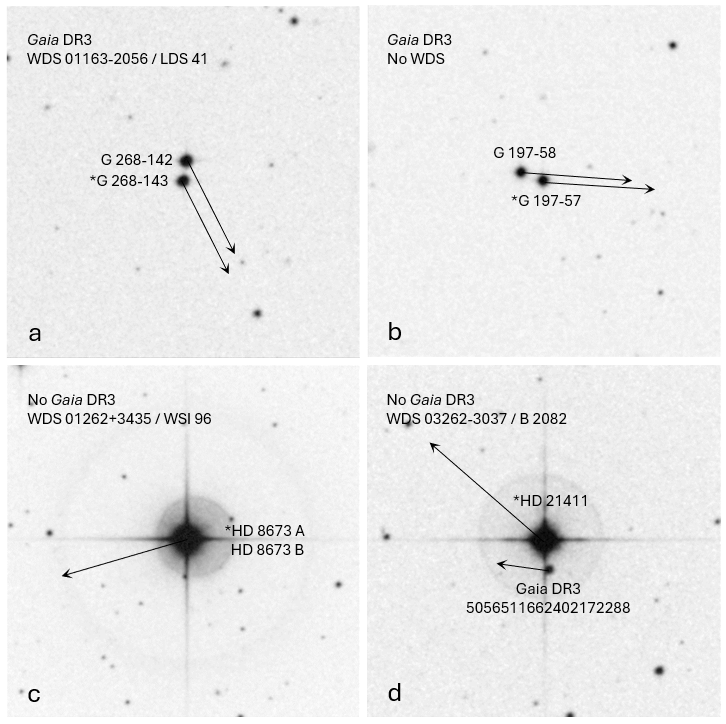}
 \caption{European Southern Observatory Digital Sky Survey DDS2 Red images centred on four representative examples of analysed systems.
 The field of view is $5 \times 5$\,arcmin, north is up and east is left.
 Stars are labelled, and exoplanet host names are preceded by an asterisk.
 Arrows indicate modulus and direction of \textit{Gaia} proper motions.
 The four representative types are:
 (a) a resolved physical double identified in our \textit{Gaia} search and with a WDS entry,
 (b) a resolved physical double found with \textit{Gaia} and not in WDS (but reported by \citealt{gaiacollaboration21b}),
 (c) a close physical double with a WDS entry ($\rho =$ 0.31\,arcsec) but not resolved by \textit{Gaia},
 and (d) an optical double in WDS but not identified in our \textit{Gaia} search.
 }
 \label{fig:pairs_examples}
\end{figure}

\begin{figure}
 \centering
 \includegraphics[width=1\linewidth, angle=0]{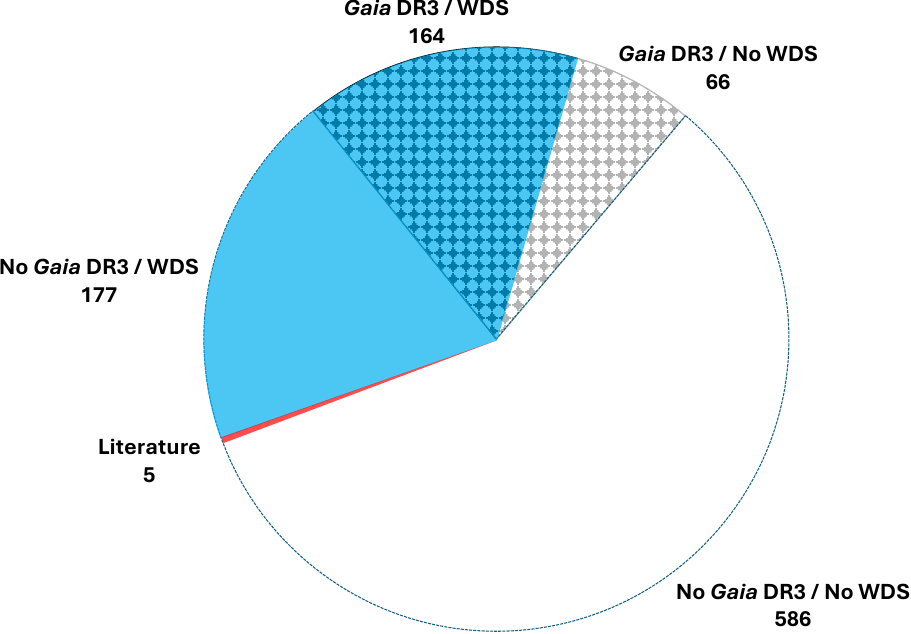}
 \caption{Pie chart of \textit{Gaia} (rhombus checkered), WDS (blue), and literature (red) systems.
 Slices are labelled with the number of systems of each type.}
 \label{fig:pie_chart}
\end{figure}

Not all the WDS pairs were identified in the \textit{Gaia} DR3 search, nor all the \textit{Gaia} DR3 common $\mu$ and $\pi$ companions were tabulated by WDS.
Most, if not all, WDS companions not found in our \textit{Gaia} DR3 search are in very close systems unresolvable by \textit{Gaia}, have moderate angular separations but accompany very bright stars, or are very faint ultracool dwarfs beyond the \textit{Gaia} limit at about $G \approx$ 20.3\,mag \citep{smart17}. 
Likewise, there are known WDS systems with additional \textit{Gaia} DR3 common $\mu$ and $\pi$ companions reported here for the first time (see below). 
As a result of our searches, we found companions to 311 of the 998 stars in the sample, including those in the 270 WDS pairs. Four representative examples of analysed systems are shown in Fig.~\ref{fig:pairs_examples}. 

We categorised the remaining 687 exoplanet host stars as single, either because WDS and the literature do not report any companions for them or because we were not able to find any common parallax and proper motion companion (by chance, there were also 687 WDS optical and physical pairs).
The list of single stars may actually be shorter, as they may have faint companions beyond the \textit{Gaia} magnitude limit, perhaps in the substellar domain \citep{smart17, smart19, marocco20}, or may be unidentified very close spectroscopic or astrometric binaries.

Finally, we also added five additional multiple systems with extremely close separations: 
the 3 circumbinary systems at less than 100\,pc (RR~Cae~AB, Kepler-16~AB, and HD~202206~AB -- \citealt{qian12,doyle11,benedict17}, respectively),
1 spectroscopic binary (HD~42936~AB -- \citealt{barnes20}),
and 1 very close pair (LP~413--32~AB -- \citealt{feinstein19}). 
The five of them have been reported in the literature but not tabulated by WDS.
We recovered them from an exhaustive object-by-object search thorough the literature. 
Fig.~\ref{fig:pie_chart} summarises the casuistry of the different types of systems in the \textit{Gaia}, WDS, and literature searches.

\subsection{Wide pairs in open clusters and associations}
\label{sec:wide_pairs_in_open_clusters_and_associations}

There are reasonable doubts about the true binding of ultra-wide pairs of young stars, as they may actually be part of stellar kinematic groups and associations \citep[][and references therein]{caballero10}.
The doubts are more reasoned when the pairs belong to nearby open clusters and OB associations, and especially when the measured separation resembles the typical separation between cluster members.
This is the case of 7 of the 311 host stars with \textit{Gaia} DR3 or WDS companions, shown in Table~\ref{tab:clusters}, which we excluded from our analysis.
They belong to the nearby Hyades open cluster \citep{reid93,perryman98} and Lower Centaurus Crux OB association \citep{dezeeuw99}. 
As expected, all of them have a large number of common $\mu$ and $\pi$ companions (and companions to companions) at wide projected physical separations, of up to 37 in the case of one of the Hyades (being the 37 companions known Hyades, too).
The stars in Table~\ref{tab:clusters} compose a complete volume-limited sample of exoplanet host stars in open clusters and dense OB associations, and could be used for chemical tagging studies linked to exoplanets \citep[e.g.][]{desilva06,liu19}.

Table~\ref{tab:clusters} does not include two close double candidates.
We kept $\epsilon$~Tau as a single host star in the Hyades although WDS tabulates a very close companion (WSI~53~Ab) undetected by \textit{Gaia}.
For this pair, \citet{mason11} measured $\rho$ = 0.237\,arcsec and $\theta$ = 108.3\,deg on only one epoch (J2005.8687), and estimated $\Delta$ = 2.4\,mag from their speckle interferometric measurements in the optical.
However, the companion candidate was not detected in an earlier speckle survey of the Hyades by \citet{mason93}, nor has been detected yet by any other team.
In particular, there are no hints for any close companion beyond the 7.6\,M$_{\rm Jup}$-minimum-mass planet around $\epsilon$~Tau \citep{sato07}. 
The other close double is b~Cen~(AB), which is made of a B2.5\,V star in the Upper Centaurus-Lupus OB association with a very close, later companion of uncertain properties and a wide-orbit substellar object detected in direct imaging \citep[][]{janson21}; this system was, however, filtered in the following~step.

\begin{table}
 \centering
 \caption[]{Host stars in open clusters and OB associations with discarded wide common $\mu$ and $\pi$ companion candidates.}
 \scalebox{0.85}[0.85]{
 \begin{tabular}{lcccc}
 \hline \hline
 \noalign{\smallskip}
 Star$^{(a)}$ & $\alpha$\,(J2000) & $\delta$\,(J2000) & $d$ & Open \\
 & (hh:mm:ss.ss) & (dd:mm:ss.s) & (pc) & cluster$^{(b)}$ \\
 \noalign{\smallskip}
 \hline
 \noalign{\smallskip} 
  HD 285507 & 04:07:01.23 & +15:20:06.1 & 45.0 & Hya  \\
  K2--25  & 04:13:05.61 & +15:14:52.0 & 44.7 & Hya \\
  K2--136 & 04:29:38.99 & +22:52:57.8 & 58.9 & Hya \\
  HD 28736 & 04:32:04.81 & +05:24:36.1 & 43.3 & Hya \\
  HD 283869 & 04:47:41.80 & +26:09:00.8 & 47.4 & Hya \\
  HD 95086 & 10:57:03.02 & --68:40:02.4 & 86.5 & LCC \\
  HD 114082 & 13:09:16.19 & --60:18:30.1 & 95.1 & LCC \\ 
 \noalign{\smallskip}
 \hline
 \end{tabular}
 }
 \label{tab:clusters} 
   \tablefoot{
    \tablefoottext{a}{The list does not contain $\epsilon$ Tau in the Hyades open cluster and b~Cen~(AB) in the Upper Centaurus-Lupus OB association.} 
    \tablefoottext{b}{Hya: Hyades open cluster; LCC: Lower Centaurus Crux OB association.} 
    }
\end{table}

\subsection{Ultracool dwarfs}
\label{sec:ultracool_dwarfs}

In their desire to be as complete as possible, either the Extrasolar Planets Encyclopaedia, the NASA Exoplanet Archive, or both often tabulate exoplanet candidates that are far from being actual exoplanets according to the International Astronomical Union definition of a planet in the Solar System\footnote{\url{https://www.iau.org/static/resolutions/Resolution_GA26-5-6.pdf}}.
For example, although much has been written about the differences between brown dwarfs and substellar objects below the deuterium burning mass limit at about 13\,M$_{\rm Jup}$ \citep[][and references therein]{caballero18}, the Extrasolar Planets Encyclopaedia tabulates the first unambiguous brown dwarfs discovered, namely Teide~1 \citep[$\sim$65\,M$_{\rm Jup}$;][]{rebolo95} and GJ~229\,B \citep[$\sim$35\,M$_{\rm Jup}$;][]{nakajima95}, counting as exoplanets\footnote{\url{https://exoplanet.eu/catalog/\#inclusion-criteria-section}}, but not hundreds of other ultracool dwarfs (UCDs) with spectral type M7\,V or later, \textit{Gaia} parallax, and mass at or below the hydrogen burning limit (e.g. \citealt{basri00,kirkpatrick05,smart19}).
Something similar happens to a few M-type companions to young stars in stellar kinematic groups and star-forming regions, and LTY-type companions to very nearby stars and brown dwarfs detected via direct imaging. 
These UCD companions, which are counted as exoplanets in one or the two catalogues, orbit at a few arcseconds from their primaries, comparable to the separations between stars in multiple systems, have been resolved from their stars and characterised photometrically and even spectroscopically, or have high-uncertainty model-dependent masses at or above the deuterium burning limit. 
Because of their extreme heterogeneity, we discarded a total of 66 stars in systems with UCDs discovered by direct imaging, which left us only exoplanet candidates in compact orbits detected with the RV and transit methods. % including b~Cen~(AB),
The 66 stars with imaged UCD companions are collected in Table~\ref{tab:UCD_systems} with their respective references.
Actually, there are 65 systems because both young late M dwarfs 2MASS J14504216--7841413 and 2MASS J14504113--7841383, which form a common $\mu$ and $\pi$ double, are each counted as exoplanet candidates.
Among the 65 discarded systems, one can find cornerstone star-brown dwarf systems such as 
GJ~229 (HD 42581) itself \citep{nakajima95},
G~196--3 \citep{rebolo98},
$\eta$~Tel \citep{lowrance00},
TWA~5 (CD--33~7795) \citep{macintosh01},
or CD--52~381 \citep{neuhauser03}.
We included here one of the three found systems with possible circumbinary planets, HD~202206, because the planet is also a brown dwarf \citep{benedict17}. 
Most of the 66 discarded stars in 65 systems in Table~\ref{tab:UCD_systems} have UCDs companions that are too faint (and, sometimes, too close) for \textit{Gaia}.

\subsection{Close binaries without \textit{Gaia} astrometric solution}
\label{close_binaries_without_Gaia}

At this stage, we were back to the 238 multiple systems that remained from the original sample of 311 stars with companions after discarding the 7 stars in Table~\ref{tab:clusters} and the 66 stars in Table~\ref{tab:UCD_systems}.
All the 238 exoplanet host stars have both $\mu$ and $\pi$ from \textit{Gaia} DR3 except for 3 very bright stars (Aldebaran, Fomalhaut, and $\gamma^{01}$~Leo), and LP~413--32~B, which was recovered (see above) from \citet{feinstein19}.
However, there are 79 companions without \textit{Gaia} DR3 $\mu$ and $\pi$.
Of them, 20 have a \textit{Gaia} DR3 entry but not a five-parameter astrometric solution:
1 star that is the early-K companion at 1.2\,arcsec to the late-F planet-host star HD~176051;
6 fainter close companions with similar or slightly greater angular separations $\rho$\,$\sim$\,1.2--5.0\,arcsec but much larger magnitude differences $\Delta G$\,$\sim$\,5.5--11.5\,mag to their primaries (Gaia DR3 959971546241605120, Gaia DR3 1220404653532465536, Gaia DR3 5917231492302779776, Gaia DR3 4122133830617000320, Gaia DR3 4296208099223616640, and Gaia DR3 6407428994689762560); 
12 stars that are wide companions to planet host stars and that are in turn part of close binaries tabulated by WDS ($\rho \approx$ 0.06--1.3\,arcsec), namely $\psi^{01}$~Aqr~[BC]\footnote{The square brackets indicate that we use these designations for the first time.}, 94~Cet~[BC] (both with very large uncertainties in \textit{Gaia} equatorial coordinates for its magnitude), HD~178911~[AB], HD~222582~[BC] (LP~703--44~AB), BD--17~588[BC], and LSPM~J1301+6337~[AB] (a wide companion to HD~113337 with a bimodal distribution in its $G$-band light curve that depends on the scan direction across the source); and 1 wide companion that is also a close binary, namely L~72--1~[AB], but is presented here for the first time (WDS 15154--7032; Sect.~\ref{sec:remarkable_systems}).
The \textit{Gaia} DR3 $G$-band light curve of L~72--1~[AB] displays the same bimodal distribution as LSPM~J1301+6337~[AB], which, together with its missing \textit{Gaia} astrometric solution, points towards a close binarity of the order of 0.2\,arcsec.

The other 59 companions do not even have a \textit{Gaia} DR3 entry.
They include the 3 companions reported in the literature but not tabulated by WDS that were mentioned above (in 2 circumbinary systems and 1 spectroscopic binary), and  56 companions in very close WDS systems, usually resolved with adaptive optics and beyond \textit{Gaia}'s capabilities (e.g. $\gamma$~Cep~AB; \citealt{neuhauser07}), that are close to naked-eye stars with very bright haloes ($\tau$~Boo~B, 26~Dra~B, 54~Psc~B), or that are very bright themselves (e.g. $\alpha$~Cen~A and~B, $\gamma^{02}$~Leo). 
Although we could recover proper motions, parallaxes, or proper motions different from \textit{Gaia} DR3 only for a small fraction of them, given the robust multiplicity classification in most cases, we kept all the known close companions in our analysis.
Thanks to this analysis, though, we were able to add a new component to a system that was considered to be double and is instead triple (WDS 15154--7032; Sect.~\ref{sec:remarkable_systems}).

\subsection{Final cut}
\label{sec:final_cut}

\begin{table*}
 \centering
 \caption[]{Systems with new common $\mu$ and $\pi$ companions.}
 \scalebox{0.83}[0.83]{
 \begin{tabular}{l@{\hspace{2mm}}l@{\hspace{2mm}}l@{\hspace{2mm}}c@{\hspace{1mm}}c@{\hspace{2mm}}c@{\hspace{2mm}}c@{\hspace{1mm}}l@{\hspace{2mm}}l@{\hspace{2mm}}c@{\hspace{2mm}}c}
 \hline \hline
 \noalign{\smallskip}
 Star & \multicolumn{2}{c}{Spectral type} & \multicolumn{2}{c}{$M_\star$} & $V_r$ & $\rho$ & WDS & Disc. code / & System & |U$_g^*$| \\
  & Value & Ref.$^{(a)}$ & (M$_{\odot}$) & Ref.$^{(a)}$ & (km~s$^{-1}$) & (arcsec)  & & Reference$^{\,(a,b)}$ & schema$^{\,(c)}$ & (10$^{33}$\,J) \\
 \noalign{\smallskip}
 \hline
 \noalign{\smallskip} 
* HD 1466$^{(d)}$ & F8\,V & Tor06 & 1.16 & Des21 & +6.53$\pm$0.16  &  & ... &  & \multirow{4}{*}{\includegraphics[width=7.5mm]{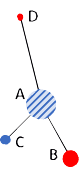}} & \multirow{4}{*}{...} \\
UCAC3 52--533 & M1.8\,(V) & Kra14 & 0.45 & Pec13 &  +5.97$\pm$24.0 & 3023 &  & This work &  & \\
UCAC3 53--724 & M5.5\,V & Kir11 & 0.12 & Pec13 & ...  & 1817 &  & Gai21 &  & \\
2MASS J00191296--6226005 & L1 & Gag15 & 0.08 & Pec13 & ... & 3772 &  & This work &  & \\
 \noalign{\smallskip}
 \hline
 \noalign{\smallskip} 
* HD 27196 & K1--2\,V & Ker22 & 0.90 & Ker22 & +0.41$\pm$0.14 &  & ... &  & \multirow{2}{*}{\includegraphics[width=6mm]{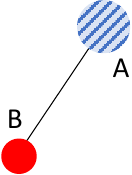}} & \multirow{2}{*}{182} \\
2MASS J04164536--2646205 & $\sim$M4.5\,V & This work & 0.18 & Pec13 & ... & 28 &  & This work &  & \\
\noalign{\smallskip}
\hline
\noalign{\smallskip}
* HD 30177 & G8\,V & Hou75 & 0.98 & Fen22 & +62.60$\pm$0.12 &  & ... &  & \multirow{2}{*}{\includegraphics[width=4.5mm]{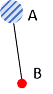}} & \multirow{2}{*}{3.8} \\
Gaia DR3 4774292312023378816 & M7\,(V) & Rey18 & 0.10 & Pec13 & ... & 780 &  & This work & \\
 \noalign{\smallskip}
 \hline
 \noalign{\smallskip} 
* HD 88072[A] & G3\,V & Gra03 & 1.04 & Fen22 & --18.00$\pm$0.13 &  & ... &  &  \multirow{3}{*}{\includegraphics[width=7.5mm]{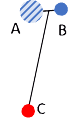}} & \multirow{3}{*}{1.3}  \\
HD 88072B & $\sim$M5\,V  & This work & 0.14 & Pec13 & ... & 3.7 &  &  Gai21 &  & \\
LP 609--39 & $\sim$M5.5\,V & This work & 0.12 & Pec13 & ... &  4772 &  & This work &  & \\
 \noalign{\smallskip}
 \hline
 \noalign{\smallskip} 
* HD 93396 & G8/K0\,IV & Hou99 & 1.46 & Tur15 & +35.00$\pm$0.12 &  & ... &  & \multirow{2}{*}{\includegraphics[width=10mm]{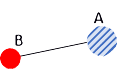}} & \multirow{2}{*}{14} \\
UCAC3 162--116679 & $\sim$K6\,V & This work & 0.69 & Pec13 & +34.60$\pm$0.71 & 1266 &  & This work &  &  \\
 \noalign{\smallskip}
 \hline
 \noalign{\smallskip} 
* HD 94834 & K1\,IV & Egg60 & 1.11 & Luh19 & +2.79$\pm$0.12 &  & ... &  &  \multirow{2}{*}{\includegraphics[width=4.5mm]{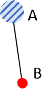}} & \multirow{2}{*}{1150} \\
Gaia DR3 3995918691799173376 & $\sim$M4\,V  & This work & 0.23 & Pec13 & ... & 3.9 &  & This work &  &  \\
 \noalign{\smallskip}
 \hline
 \noalign{\smallskip} 
* HD 96700 & G0\,V & Gra06 & 0.99 & Tur15 & +12.80$\pm$0.12 &  & ... &  & \multirow{2}{*}{\includegraphics[width=4.5mm]{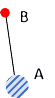}} & \multirow{2}{*}{6.7} \\
CD--27 7881 & K6\,V & Gra06 & 0.72 & Pec13 & ... & 6867 &  & This work &  & \\
 \noalign{\smallskip}
 \hline
 \noalign{\smallskip} 
* CD--39 7945 & $\sim$G8\,V & This work & 0.92 & Fri20 & --14.00$\pm$0.16 &  & ... & & \multirow{2}{*}{\includegraphics[width=8mm]{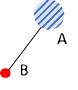}} & \multirow{2}{*}{2.9} \\
Gaia DR3 6140359952471910272 & $\sim$M5.5\,V & This work & 0.12 & Pec13 & ... & 726 &  & This work & &  \\
 \noalign{\smallskip}
 \hline
 \noalign{\smallskip} 
* UPM J1349--4603 & $\sim$M3\,V & This work & 0.39 & Hob23 & --15.00$\pm$2.72 &  & ... &  & \multirow{2}{*}{\includegraphics[width=12mm]{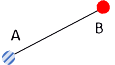}} & \multirow{2}{*}{4.7} \\
Gaia DR3 6107347154502415744 & D: & Gen19 & 0.65 & Gen19 & ... & 1361 &  & This work &  &  \\
 \noalign{\smallskip}
 \hline
 \noalign{\smallskip} 
* HD 134606 & G6\,IV & Gra06 & 1.06 & Luh19 & +1.94$\pm$0.12 &  & 15154--7032 &  & \multirow{2}{*}{\includegraphics[width=8mm]{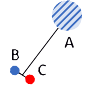}} & \multirow{2}{*}{1194} \\
L 72--1[AB] & $\sim$M3+$\sim$M3 & This work & 0.74 & Pec13 & +8.54$\pm$6.11 & 57 &  & FMR 173/This work &  & \\
 \noalign{\smallskip}
 \hline
 \noalign{\smallskip} 
* HD 135872 & G5\,IV & Hou99 & 1.65 & Fen22 & --20.00$\pm$0.13 &  & ... &  & \multirow{2}{*}{\includegraphics[width=5.5mm]{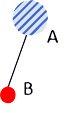}} & \multirow{2}{*}{61} \\
Gaia DR3 6334935890969037696 & $\sim$K4\,V & This work & 0.73 & Pec13 & --20.00$\pm$0.48 & 423 &  & This work & \\
 \noalign{\smallskip}
 \hline
 \noalign{\smallskip} 
* CD--24 12030 & $\sim$K2.5\,V & This work & 0.80 & Chr22 & +5.17$\pm$0.29 &  & ... & & \multirow{3}{*}{\includegraphics[width=16mm]{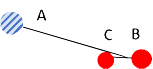}} & \multirow{3}{*}{51} \\
Gaia DR3 6214879765359934592 & $\sim$M2\,V & This work & 0.45 & Pec13 & +8.78$\pm$3.18 & 231 &  & This work &  &  \\
Gaia DR3 6214879769657777536 & $\sim$M4\,V & This work & 0.25 & Pec13 & +42.30$\pm$4.30 & 226 &  & This work &  &  \\
 \noalign{\smallskip}
 \hline
 \noalign{\smallskip} 
* HD 143361 & G6\,V & Hou78 & 0.95 & Min09 & --0.60$\pm$0.16 &  & ... &  & \multirow{2}{*}{\includegraphics[width=15mm]{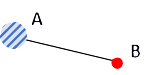}} & \multirow{2}{*}{2.3} \\
Gaia DR3 5994315469412318848 & $\sim$M4.5\,V & This work & 0.19 & Pec13 & ... & 1964 &  & This work & & \\
 \noalign{\smallskip}
 \hline
 \noalign{\smallskip} 
* TOI--1410 & $\sim$K4\,V & This work & 0.71 & Sta19 & +2.46$\pm$0.10 &  & ... & & \multirow{2}{*}{\includegraphics[width=12mm]{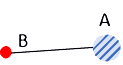}} & \multirow{2}{*}{68} \\
Gaia DR3 1958584427911285120 & $\sim$M3\,V & This work & 0.36 & Pec13 & +3.38$\pm$1.87 & 91 &  & This work & & \\
 \noalign{\smallskip}
 \hline
 \noalign{\smallskip} 
* HD 222259A & G6\,V & Tor06 & 0.96 & Ben19 & +7.13$\pm$0.28 &  & 23397--6912 &  & \multirow{3}{*}{\includegraphics[width=16mm]{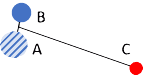}}  & \multirow{3}{*}{4.6} \\
HD 222259B & K3\,V & Tor06 & 0.78 & Pec13 & +5.17$\pm$0.42 & 5.4 &  & R   348 &  & \\
2MASS J23321028--6926537 & M5.3\,V & Ujj20 & 0.17 & Pec13 & ... & 2560 &  & This work &  & \\
 \noalign{\smallskip}
 \hline
 \end{tabular}
 } % \scalebox
 \label{tab:new_detected_systems}  
\tablefoot{
    \tablefoottext{a}{References: 
      Alo15: \citet{alonsofloriano15}; 
      Ben19: \citet{benatti19}; 
      Chr22: \citet{christiansen22}; 
      Des21: \citet{desidera21};
      Egg60: \citet{eggen60}; 
      Fen22: \citet{feng22}; 
      Fri20: \citet{fridlund20}; 
      Gag15: \citet{gagne15}; 
      Gai21: \citet{gaiacollaboration21b}; 
      Gen19: \citet{gentilefusillo19}; 
      Gra03: \citet{gray03}; 
      Gra06: \citet{gray06}; 
      Hob23: \citet{hobson23}; 
      Hou75: \citet{houk75}; 
      Hou78: \citet{houk78}; 
      Hou99: \citet{houk99}; 
      Ker22: \citet{kervella22};
      Kir11: \citet{kirkpatrick11}; 
      Kra14: \citet{kraus14}; 
      Lep13: \citet{lepine13}; 
      Low00: \citet{lowrance00};
      Luh19: \citet{luhn19}; 
      Min09: \citet{minniti09}; 
      Pec13: \citet{pecaut13}; 
      Rey18: \citet{reyle18}; 
      Rie17: \citet{riedel17b}; 
      Smi18: \citet{smith18}; 
      Sta19: \citet{stassun19}; 
      Tor06: \citet{torres06}; 
      Tur15: \citet{turnbull15};
      Ujj20: \citet{ujjwal20}.}
    \tablefoottext{b}{WDS discoverer codes are written in uppercase, and literature references in lowercase.} 
    \tablefoottext{c}{Blue circles: already reported stars. Red circles: new stars in system. Stripped circles: exoplanet host stars. Symbol sizes are proportional to stellar masses.}
    \tablefoottext{d}{The HD~1466 system is in the Tucana-Horologium association; we only tabulate \textit{Gaia} DR3 companions at $s <$ 1\,pc.}
}
\end{table*}

\begin{table*}
 \centering
 \caption[]{Binary systems with detected planets around both stars.}
  \scalebox{0.85}[0.85]{
 \begin{tabular}{lcccccccc}
 \hline \hline
 \noalign{\smallskip}
 Star & \multicolumn{2}{c}{Spectral type} & $\alpha$\,(J2000) & $\delta$\,(J2000) & $d$ & \multicolumn{2}{c}{Planets}  & \textit{s}\\
  & Value & Reference$^{(a)}$ & (hh:mm:ss.ss) & (dd:mm:ss.s) & (pc) & Number & Reference$^{(a)}$ & (au) \\
 \noalign{\smallskip}
 \hline
 \noalign{\smallskip} 
 HD 20782 & G1.5\,V & Gra06 & 03:20:03.58 & --28:51:14.7 & 35.9 & 1 & Jon06 & \multirow{2}{*}{9\,075} \\
 HD 20781 & G9.5\,V & Gra06 & 03:20:02.94 & --28:47:01.8 & 36.0 & 4 &  May11, Udr19 & \\
 \noalign{\smallskip}
 UCAC3 234--106607 & M1.3\,V & Bir20 & 12:58:23.27 & +26:30:09.1 & 58.7 & 1 & Lew22 & \multirow{2}{*}{909} \\
 Sand 178$^{(b)}$ & M2.4\,V & Bir20 & 12:58:22.24 & +26:30:16.1 & 58.8 & 1 & Lew22 & \\
 \noalign{\smallskip}
 HD\,133131A & G2.0\,V & Sto72 & 15:03:35.45 & --27:50:33.2 & 51.5& 2 & Tes16 & \multirow{2}{*}{379} \\
 HD\,133131B & G2.0\,V & Sto72 & 15:03:35.81& --27:50:27.6 & 51.5 & 1 & Tes16 & \\
 \noalign{\smallskip}
 \hline
 \end{tabular}
 }
 \label{tab:both_stars_with_planets}
   \tablefoot{
    \tablefoottext{a}{
    Bir20: \citet{birky20}; 
    Gra06: \citet{gray06}; 
    Jon06: \citet{jones06}; 
    Lew22: \citet{lewis22}; 
    May11: \citet{mayor11}; 
    Sto72: \citet{stock72}; 
    Tes16: \citet{teske16}; 
    Udr19: \citet{udry19}.}
    \tablefoottext{b}{\citet{lewis22} stated that ``depending on the true period and eccentricity of the system, the minimum companion mass may fall in a range from 2\,M$_{\rm Jup}$ to 50\,M$_{\rm Jup}$, making it a possible brown dwarf candidate''.}
   }
\end{table*}

We went on with the revision of known exoplanet candidates, and discarded 18 additional systems from our analysis:
1 system, namely GJ~682, with a bright companion candidate at 0.17\,arcsec reported by \citet{ward15} and discarded with deep imaging by \citet{desgrange23};
2 systems, namely 14~Her and 70~Vir, with faint companion candidates ($\Delta I=$ 10.9--11.4\,mag, $\rho=$ 2.9--4.3\,arcsec and 42.8\,arcsec) detected in imaging searches by \citet{pinfield06} and \citet{roberts11} on only one epoch that are background sources according to, e.g., \citet{patience02b}, \citet{grether06}, \citet{carson09}, \citet{leconte10}, \citet{ginski12}, \citet{durkan16}, and \citet{fontanive21} (see also \citealt{rodigas11} and \citealt{dalba21} for searches at even closer angular separations);
2 systems, namely HD~9578 and TOI--717, with planet candidates that have not undergone publication within a peer-reviewed journal yet\footnote{HD~9578 was only announced in a press release in October 2009 during an international conference.};
1 system, namely Fomalhaut ($\alpha$~PsA), with a planet candidate proposed by \citet{kalas05, kalas08}, considered to be the first candidate imaged at visible wavelengths, which is actually an expanding blob of debris from a massive planetesimal collision in the disc \citep{gaspar20, gaspar23};
1 triple system, namely LP~563--38, with two M dwarfs and an eclipsing brown dwarf \citep[][]{irwin10}; 
6 systems, namely 11~Com,
HD~26161,
HD~77065,
HD~109988,
HD~127506,
and BD+24~4697, with radial-velocity companions with minimum masses between 15.5\,M$_{\rm Jup}$ and 53.0\,M$_{\rm Jup}$ and, therefore, actual masses well above the deuterium burning mass limit; % \citep[$M < 500\,M_{\rm Jup}$;][]{wilson16}.
and 5 systems with exoplanet candidates with astrometric masses above the hydrogen burning mass limit, i.e. in the stellar domain.
The 5 discarded stellar companions are:
HD~184601\,B \citep[$M = 117^{+36}_{-32}\,{\rm M}_{\rm Jup}$;][]{xiao23}, 
HD~211847\,B \citep[$M = 148\pm5\,{\rm M}_{\rm Jup}$;][]{philipot23}, 
HD~283668\,B \citep[$M = 319\pm19\,{\rm M}_{\rm Jup}$;][]{xiao23}, 
HD~114762\,B \citep[$M = 210\pm10\,{\rm M}_{\rm Jup}$;][]{gaiacollaboration23}, 
and BD--02~2198\,B ($M = 196.9^{+5.0}_{-4.9}\,{\rm M}_{\rm Jup}$; \citealt{baroch21} -- see also \citealt{biller22}).
Of them, HD~114762\,B has received much attraction in the last decades (\citealt{latham89,patience02b,bowler09,kane11,kiefer19} -- See \citealt{latham12} for a historical review).
On the contrary to the 66 discarded stars in Sect.~\ref{sec:ultracool_dwarfs}, these 18 additional systems have companion candidates that have never been resolved from their host stars.

Including the new companion in the close binary with missing \textit{Gaia} DR3 astrometric solution and bimodal distribution of $G$-band light curve (L~72--1~[AB]), we had a total of 64 % 44 known + 3 no RV + 17 new
companions found in our common $\mu$ and $\pi$ search but not reported by WDS.
We carefully investigated the literature associated with these 64 stars and found that 44 of them had already been proposed as companions to the exoplanet host stars by other authors \cite[e.g.][]{fontanive21,gaiacollaboration21b}.
We looked for available \textit{Gaia} (mean) absolute radial velocities for the remaining 20 stars (in 18 systems).
Because of the typical faintness of the new companions reported here and the \textit{Gaia} limit at $G \sim$ 16\,mag for radial velocities \citep{katz23}, there are measurements for both the exoplanet host star and companion candidates for only 8 systems.
RAVE and other radial-velocity surveys \citep{steinmetz06,kunder17} did not provide with additional measurements of companions.
Of the 8 systems, 3 binary systems, namely those with exoplanet host stars HD~168009, HD~210193, and K2--137, have radial velocities that differ by $80.90\pm0.27$, $32.55\pm0.19$, and $34.4\pm2.9$\,km\,s$^{-1}$, respectively.
Such differences may be difficult to ascribe to unknown close companions to the secondaries, as it may actually be happening at lesser degree to the M-dwarf companion to CD--24~12030 or the G--K components of HD~222259.
Therefore, the stars in the three pairs do not have the same galactocentric velocities and are optical pairs.
Hence, we discarded them from our analysis.
This deletion left us with 17 genuinely new companions candidates, which are discussed in Sect.~\ref{sec:new_stellar_systems} and shown in Tables~\ref{tab:new_detected_systems} and~\ref{tab:new_detected_systems_astrometry}.
The former tabulates main derived data (stellar mass $M_\star$, systemic radial velocity $V_r$, angular separation $\rho$, WDS identifier and discoverer code when available, a pictographical system schema, and reduced binding energy $|U_g^*|$ -- see below), while the later tabulates the astrometric data (parallax, proper motion in right ascension and declination, $\mu_{\rm ratio}$, and $\Delta$PA).

After all these considerations, we identified 215 exoplanet host stars in 212 multiple systems, of which 173 are binary, 39 are triple, and 3 are quadruple. 
The list of these multiple systems is shown in Table~\ref{tab:systems1}.
We tabulate only 212 entries because there are 3 binary systems with planets discovered around both stars.
The 3 systems, accounting for 6 stars and 10 planets, are also displayed in Table~\ref{tab:both_stars_with_planets}. 

\subsection{Planetary systems}
\label{sec:results_planetary_systems}

\begin{figure*}
 \centering
 \includegraphics[width=0.9
\linewidth, angle=0]{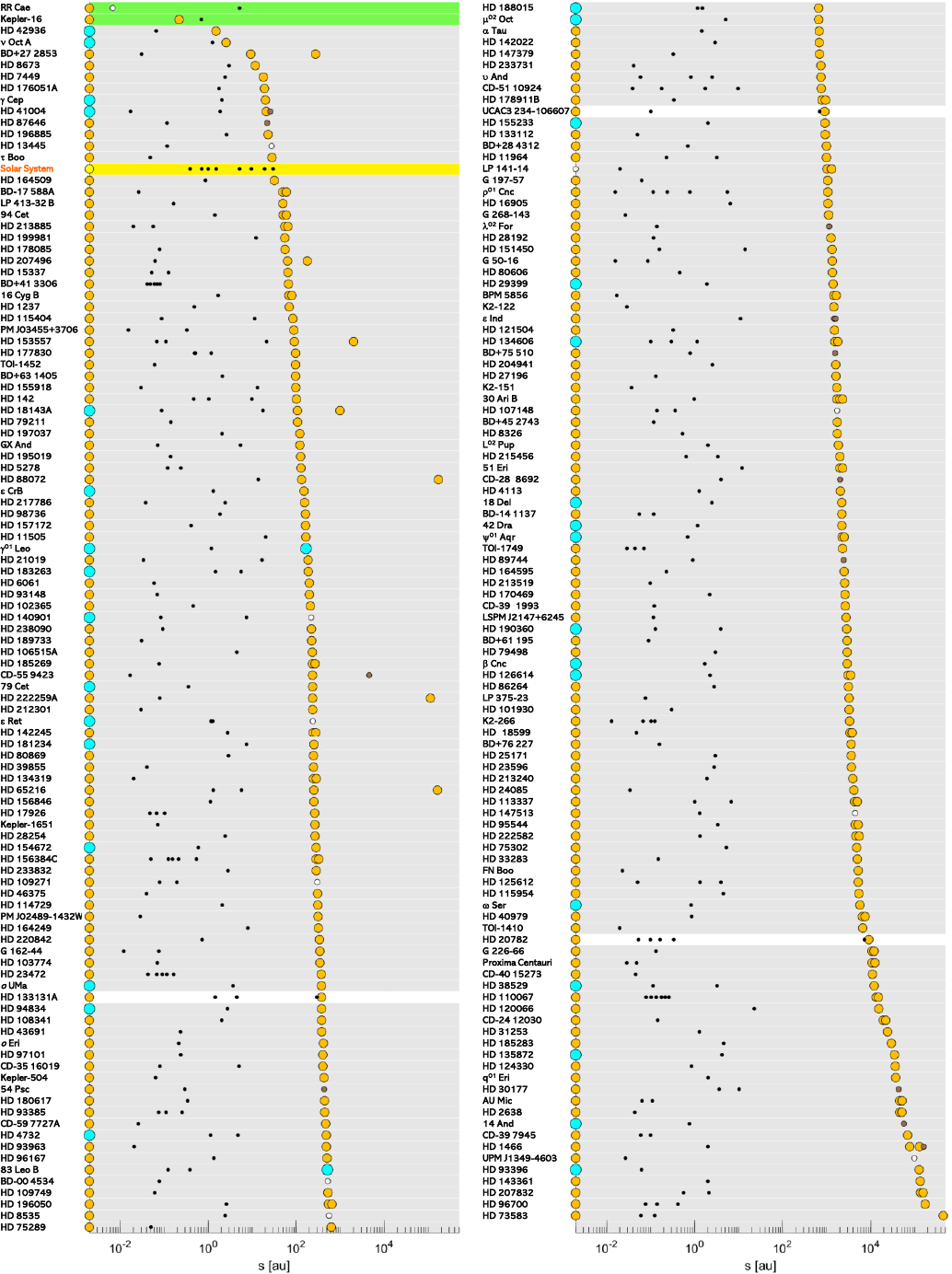}
 \caption{Schematic configurations of multiple stellar systems with exoplanets.
 Orange circles are main-sequence stars, cyan circles are subgiant and giant stars, white circles are white dwarfs, small brown circles are brown dwarfs, and black dots are planets.
 We display our 212 systems with grey background, except for the two systems with circumbinary planets, namely RR~Cae and Kepler-16, with green background, and the three systems from Table~\ref{tab:both_stars_with_planets} with planets around both stars, with white background.
 The systems are sorted by increasing separation from the planet host star to the closest companion star.
 The abscissa is in logarithmic scale.
 We also display the Solar System in yellow as a comparison.}
 \label{fig:planets}
\end{figure*}

For all the 215 stars in 212 multiple systems, we compiled the main parameters of 276 of the 302 planets orbiting them from the Extrasolar Planets Encyclopaedia and NASA Exoplanet Archive. 
In particular, we retrieved their orbital periods, $P$, semi-major axes, $a$, eccentricities, $e$, and masses, $M_{\rm pl}$, or minimum masses, $M_{\rm pl} \sin{i}$, for transiting or radial-velocity planets, respectively.
When exoplanets are common to both databases, we took the most recent parameters from the NASA Exoplanet Archive, except for very few cases with significantly smaller uncertainties that we took from the Extrasolar Planets Encyclopaedia.
The remaining 26 planets miss at least one datum.
The schematic configurations of the 212 systems with all their members (stars, white dwarfs, brown dwarfs, planets) are represented in Fig.~\ref{fig:planets}.
In occasions, the planet host star is not the brightest primary in the system (e.g. $\alpha$~Cen, 26~Dra, 83~Leo, GJ~667).
A total of 32 of the investigated exoplanets orbit giant or subgiant stars and only one around a white dwarf (LP~141--14\,b; Sect.~\ref{sec:remarkable_systems}). 
The rest of the host stars are main-sequence stars (see Sect.~\ref{sec:discussion} and Fig.~\ref{fig:HRD_new_detected_systems}).

Most exoplanet searches have focused on stars without close stellar companions.
For example, the CARMENES survey for exoplanets around nearby M dwarfs \citep{quirrenbach14} discarded from their guaranteed and legacy time observations target stars that have companions at less than 5\,arcsec, including astrometric and spectroscopic binaries, but kept resolved stellar systems with wider separations \citep{caballero16,cortes17b,jeffers18,reiners18b}.
Since the CARMENES survey is a prototypical radial-velocity exoplanet search and our sample is built upon a collection of results of searches, our target list of 212 systems is naturally biased against very close binary systems with planets. 
Acknowledging this bias prior to commencing any discussion is, therefore, mandatory.
Furthermore, this deficit of close binary systems is also found in transit exoplanet searches, which are also partly affected by this bias \citep{ziegler21}. 

For comparison purposes, we defined a control sample of single stars with exoplanets, but no known stellar or brown dwarf companion at any separation.
For that, we took a few steps back in our sample definition and discarded multiple systems with planets (Table~\ref{tab:systems1}), or with one or more ultracool dwarfs (Table~\ref{tab:UCD_systems}), and retained only single stars at $d <$ 100\,pc.
For the 687 single stars with planets, we searched for $P$, $a$, $e$, and $M_{\rm pl}$ or $M_{\rm pl} \sin{i}$ of their 1029 planets exactly as we did for the multiple systems with planets. 
From now on, we called this set of stars the ``single star sample''.

\section{Results}
\label{sec:results_discussion}

\subsection{New stellar systems}
\label{sec:new_stellar_systems}

\begin{figure}
 \centering \includegraphics[width=1\linewidth, angle=0]{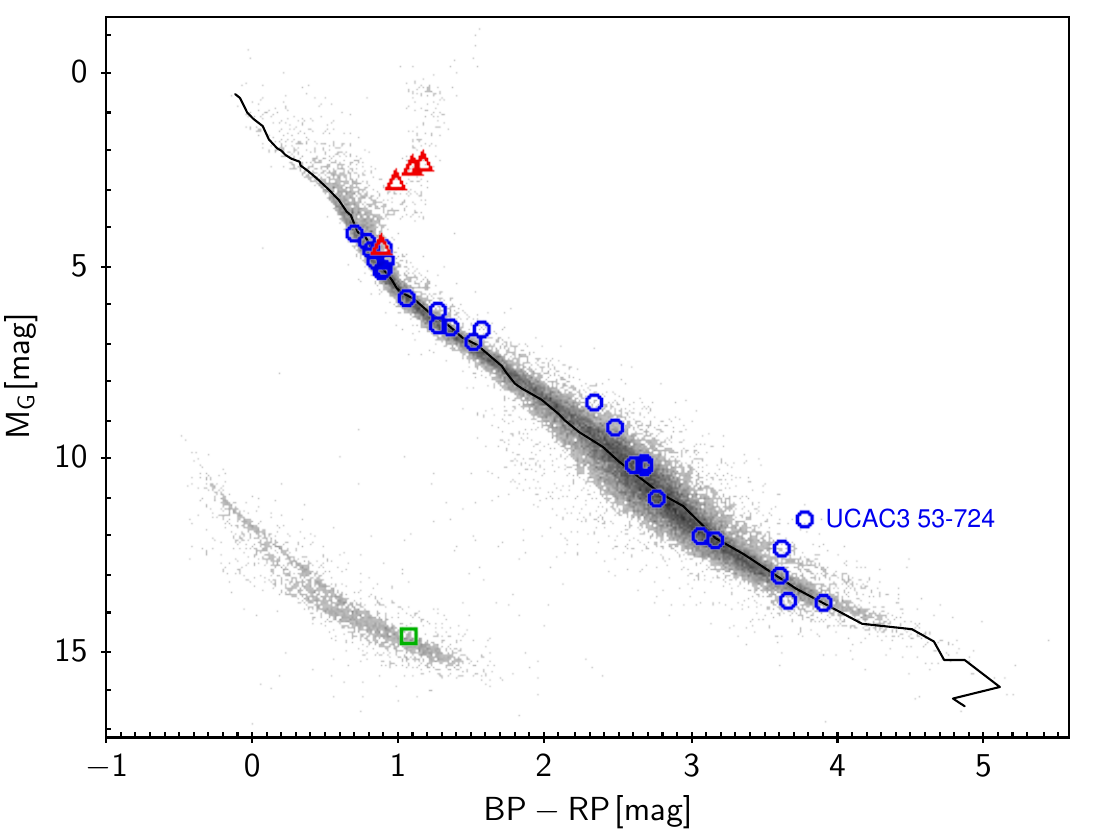}
 \caption{\textit{Gaia} colour-magnitude diagram of the systems with new common $\mu$ and $\pi$ companions in Table~\ref{tab:new_detected_systems}.
 Blue circles, red triangles, and the green square stand for the main-sequence, subgiants, and white dwarf stars, while the small grey points are over 50\,000 random \textit{Gaia} sources retrieved as \citet{taylor19} did.
 The young overluminous star UCAC3\,53--724 in Tucana-Horologium is labelled.
 The black solid line is the updated main sequence of \citet{pecaut13}.
 We did not apply any colour or magnitude correction for reddening.}
 \label{fig:HRD_new_detected_systems}
\end{figure}

Of the 17 genuinely new companions in Table~\ref{tab:new_detected_systems}, 12 are part of completely new systems, while the other 5 are additional companions to 4 systems tabulated by WDS (2) or reported in the literature (2).
In these 4 systems, the K- and M-dwarf companions had never been described in the literature and, therefore, we tabulate their \textit{Gaia} DR3 identifiers (the companion of HD~143361 was catalogued by \citealt{gaiacollaboration21b}).
On the other hand, the 17 companions are members of 15 systems, 4 of which are triple, and 1 is quadruple.
The 4 known systems with new additional companions and the new triple system are described in Sect.~\ref{sec:remarkable_systems}.
The remaining 10 systems are new binaries, which were thought to be single.

As already noticed by \citet{shaya11}, \citet{newton19}, or \citet{gaiacollaboration21b}, the very large projected physical separations between primary stars and companions casts doubts on the actual gravitational binding of some systems.
In our sample, a few of the systems are very wide. 
In particular, there are at least two systems in young associations and separations greater than 1000\,arcsec (HD~1466 with its M-L companions, and WDS 23397--6912), which make the multiple system classification even blurrier \citep{caballero10,tokovinin12,duchene13}.

To assess whether the 17 genuinely new companions in Table~\ref{tab:new_detected_systems} are indeed bound or not, we calculated the reduced binding energy value by following the methodology outlined by \citet{caballero09}:

\begin{equation}
|U^*_g| = G\frac{M_{\star,1} M_{\star,2}}{s},
\label{eqn:binding}
\end{equation}

\noindent where $G$ is the gravitational constant, $M_{\star,1}$ and $M_{\star,2}$ are the masses of both stars, and $s$ their projected physical separation.
The asterisk in $|U^*_g|$ indicates that the absolute values of the ``true'' potential energies $U_g$ using the physical separation $r$ must be lower than in Table~\ref{tab:new_detected_systems} \citep{caballero09}.
For the three triple systems, we used the combined masses of the close pairs in the calculus (e.g. at a large separation, LP~609--39 feels the gravitational attraction of HD~88072AB as if it were a single, more massive star).
We refrained from using the $\overline{r} \approx 1.26 ~ \overline{s}$ relation between projected and true separations determined by \citet{fischer92} to facilitate direct comparisons with certain other studies \citep{close03, burgasser07a, radigan09, caballero10, faherty10}.

We computed the projected physical separation from the angular separation and the (parallactic) distance to the primary and used the colour-magnitude diagram in Fig.~\ref{fig:HRD_new_detected_systems} as a guide to determine stellar masses.
For resolved stars in the main sequence, we used the conversion from \textit{Gaia} $G$-band absolute magnitudes and $B_P-R_P$ colours to masses outlined in Table~5 of \citet{pecaut13}, which is available in an enhanced and updated version on line\footnote{\url{https://www.pas.rochester.edu/~emamajek/EEM_dwarf_UBVIJHK_colors_Teff.txt}}.
For six stars out of the main sequence (e.g., four subgiants, one white dwarf, and one young overluminous star), we took their masses from the literature \citep{turnbull15, livingston18, gentilefusillo19, luhn19, feng22}.
For completeness, we also took spectral types from Simbad; when not available, we estimated them from photometry by relying on \citet{pecaut13} and \citet{cifuentes20}.

The names, spectral types (and references), stellar masses (and references), radial velocities for the 35 resolved stars, and angular separation, WDS identifier, pair discovery code or reference, and a schema of the 15 systems with new proper motion and parallax companions are displayed in  Table~\ref{tab:new_detected_systems}, and their astrometric properties are shown in Table~\ref{tab:new_detected_systems_astrometry}.
The asterisks denote the planet-host stars, which in all cases are the system primaries.
We were able to compute $|U^*_g|$ values for all but one trapezoidal system, of which the host star is HD~1466 in Tucana-Horologium \citep{gonzalezpayo23}.  
The reduced binding energies vary from over $10^{36}$\,J for the HD~94834 binary (K1\,IV + $\sim$M4\,V) to merely a few $10^{33}$\,J for the widest systems, with separations of up to 6\,900\,arcsec.
These small $|U^*_g|$ values are, however, slightly greater than $10^{33}$\,J, which may represent the minimum threshold for binding gravitational energy before disruption by the galactic potential \citep{caballero10}, as well as the minimum value of reduced binding energy of moderately separated binary systems of very low mass \citep{chauvin04,artigau07,caballero07,radigan09}.
Even for the system HD~88072\,AB and LP~609--39, the lowest $|U^*_g|$ in Table~\ref{tab:new_detected_systems} do not deviate from what is expected for gravitationally bound systems in the Milky Way (see Fig.~6 of \citealt{gonzalezpayo23}).
Besides, the actual binding of HD~1466 and its companions in Tucana-Horologium and similar systems will be the subject of a forthcoming work, and it is out of the scope of this paper.
Furthermore, we kept in Table~\ref{tab:systems1} two systems with ultra-wide separations separations greater than 6900\,arcsec: WDS~14396--6050 ($\alpha$~Cen~AB and Proxima, $\rho \sim$ 7960\,arcsec; \citealt{innes15}) and WDS~08388--1315 (HD~73583 and BD--09~2535, $\rho \sim$ 14\,240\,arcsec; %14\,236\,arcsec; 
\citealt{shaya11}).
As a result, we maintained the 15 systems for the following analysis.

\subsection{Semi-major axes and separations}
\label{sec:separations}

\begin{figure}
 \centering
 \includegraphics[width=\linewidth, angle=0]{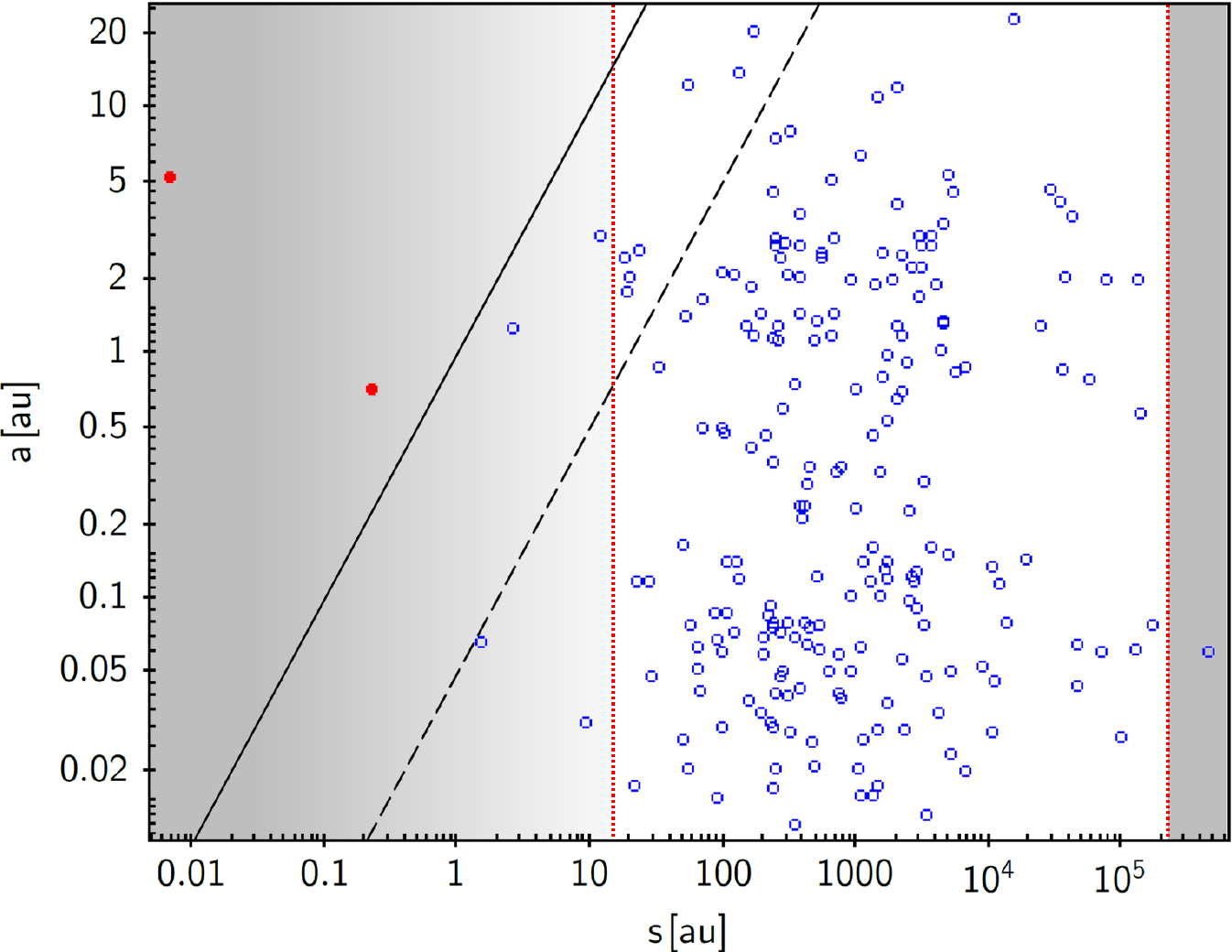}
 \caption{Exoplanet semi-major axis vs. star projected physical separation.
 Solid and dashed black lines mark the $s$:$a$ relationships at 1:1 and 20:1, respectively.
 The two circumbinary systems are marked with red filled symbols.
 Grey regions delimited by red dotted lines (16.5\,au and 1\,pc = 648\,000/$\pi \approx$ 206\,264.81\,au) indicate the approximate incompleteness areas (see main text). 
 }
 \label{fig:s-a-plot}
\end{figure}

As a first step of the analysis, we compared the projected physical separations between stars in multiple systems, $s$, and the exoplanet semi-major axes, $a$. 
In particular we compared the $s$ between the exoplanet host star and the closest stellar (or white dwarf, or brown dwarf) companion in triple and quadruple systems, and the $a$ of the most separated planet in multi-planet systems.
This comparison is illustrated by Fig.~\ref{fig:s-a-plot}.
We mark the incompleteness areas in shadows of grey.
The outer limit of the completeness area is defined by our maximum search $s$ at 1\,pc, while the approximate inner limit is set by $\overline{s} = \rho_{Gaia} \cdot \overline{d}$, where $\overline{s}$ and $\overline{d}$ are the median of the individual $s$ and $d$, and $\rho_{Gaia}$ is the critical value at 0.4\,arcsec for spatial resolution of close binaries by \textit{Gaia} \citep{lindegren18a,lindegren21}.
There have been, though, adaptive optics and speckle imaging searches that have explored inner regions ($\rho <$ 0.4\,arcsec).
Besides, the investigated range of projected physical separations is directly proportional to the $d$, so the inner regions of nearby stars are in general better studied, and vice versa. 
Therefore, the darker the region, the more incomplete it is.

In Fig.~\ref{fig:s-a-plot}, besides the two close binaries with circumbinary planets (RR~Cae and Kepler--16), there is a number of remarkable systems that stand out because of their relatively small $s/a$ ratios.
They are relatively close binaries with planetary systems, some of which might challenge current exoplanet formation scenarios in truncated protoplanetary discs.
%About 75\% of the displayed multiple systems have $s/a > 20^2 = 400$.
About 25\% of the displayed multiple systems have $s/a < 188$ (first quartile), and about 10\% have $s/a < 34$ (first decile).
We list in Table~\ref{tab:systems_with_sa20} the only 9 systems with $s/a <$ 20, together with their corresponding references.
The 9 systems represent about 8\% of the systems (there is a gap in the $s/a$ distribution at $\sim$13--22).
Most of the references, especially those presenting systems with the smallest $s/a$ ratios, already made extensive discussion on the different formation, evolution, and stability mechanisms that gave rise to such peculiar systems \citep[e.g.][]{neuhauser07,ramm21,feng22}.
Some stellar parameter values may be biased due to companion blend (e.g. HD~176051\,AB: \citealt{muterspaugh10} did not know around which component the planet b, detected by astrometry, is orbiting).
Actually, some stellar systems secondaries, marked with `[B]' in Table~\ref{tab:systems_with_sa20}, are so close to their primaries that they do not have an entry in Simbad and have only been resolved with high-resolution imagers.
The non-tabulated system that stands out in Fig.~\ref{fig:s-a-plot} with $s \approx$ 1.5\,au and $a \approx$ 0.07\,au ($s/a \approx$ 22) is HD~42936, which is made of a K0\,IV primary orbited by a super-Earth at $P \approx$ 6.67\,d and a very-low mass star at $P \approx$ 507\,d \citep{barnes20}.

According to \citet{winn15}, ``the rough rule-of-thumb [for system stability is] that the planet’s period should differ by at least a factor of three from the binary’s period, even for a mass ratio as low as 0.1''.
The three stars in Table~\ref{tab:systems_with_sa20} with the smallest $s/a$ ratios are: 
HD~199981 \citep[][$s/a \approx$ 4.48]{feng22}, % P ratio about 8
HD~8673 \citep[][$s/a \approx$ 3.97]{hartmann10,feng22}, % P ratio about 7
and $\nu$~Oct \citep[][$s/a \approx$ 2.04]{ramm21}. % P ratio about 2.6
Their corresponding $P_{\rm star} / P_{\rm planet}$ ratios, after retrieving the stellar masses from the literature \citep[e.g.][]{roberts15} and applying Third Kepler's Law, are  about 8, 7, and 2.6, respectively.
HD~199981 (a late-K primary and a mid-M secondary separated by 2.7\,arcsec) and HD~8673 (a late-F primary and an early-M secondary separated by 0.31\,arcsec), both with massive substellar companions at the planet-brown dwarf boundary, seem to be stable systems in spite of their small $s/a$ and $P_{\rm star} / P_{\rm planet}$ ratios (see again \citealt{feng22}).
However, the $\nu$~Oct system (an early K giant with a $\sim$0.58\,M$_\odot$ companion separated by 0.11\,arcsec and a planetary candidate in a retrograde orbit -- \citealt{ramm09,ramm16,ramm21,ramm15}), is catalogued in the Extrasolar Planets Encyclopaedia as ``Unconfirmed'' and is not catalogued at all by the NASA Exoplanet Archive.
As a result, the hypothetical challenge for formation and stability scenarios may not apply in this particular case, nor in the other confirmed systems with larger $s/a$ and $P_{\rm star} / P_{\rm planet}$ ratios.

\begin{table*}
 \centering
 \caption[]{Multiple systems with planets and $s/a<20$.}
  \scalebox{0.85}[0.85]{
 \begin{tabular}{l@{\hspace{1mm}}ll@{\hspace{1mm}}cccl}
 \hline \hline
 \noalign{\smallskip}
 Host star & Companion & Planet & $s$ & $a$ & $e$ & References \\  
  & star &  & (au) & (au) &  &  \\ 
 \noalign{\smallskip}
 \hline
 \noalign{\smallskip} 
 HD 11505[A] & HD 11505B & b & 167.19$\pm$0.20 & 20.2$^{+4.8}_{-3.4}$ & 0.144$^{+0.011}_{-0.056}$  & \citet{feng22} \\ 
 \noalign{\smallskip}
 HD 88072[A] & HD 88072B & b & 132.77$\pm$0.19 & 13.9$^{+4.1}_{-2.0}$ & 0.16$^{+0.14}_{-0.10}$ & \citet{feng22} \\ 
 \noalign{\smallskip}
 HD 8673[A] & HD 8673[B]  & b & 11.8 & 2.97$^{+0.15}_{-0.17}$  & 0.730$^{+0.042}_{-0.026}$  & \citet{hartmann10,feng22} \\ 
 \noalign{\smallskip}
 HD 196885 & HD 196885B & b & 23.0 & 2.6$\pm$0.1 & 0.48$\pm$0.02  & \citet{correia08,chauvin11}  \\ 
 \noalign{\smallskip}
 HD 7449[A] & HD 7449[B] & b & 20.8$\pm$0.3 & 2.438$^{+0.062}_{-0.063}$ & 0.752$^{+0.035}_{-0.032}$  & \citet{dumusque11,feng22}  \\ 
 \noalign{\smallskip}
 $\gamma$ Cep [A] & $\gamma$ Cep [B] & b & 20.18$\pm$0.66 & 2.05$\pm$0.06 & 0.049$\pm$0.034 & \citet{hatzes03,neuhauser07,endl11} \\ % Añade una nota debajo diciendo que esta es la estrella de Campmbell et al. 1989
 \noalign{\smallskip}
 HD 176051A & HD 176051B & b & 19.14$\pm$0.04 & 1.76  & 0 (fixed)  & \citet{muterspaugh10} \\ % Some stellar parameter values may be biased due to companion blend (WDS Catalog)
 \noalign{\smallskip}
 HD 199981[A] & HD 199981B & b  & 55.16$\pm$0.03 & 12.3$^{+3.6}_{-2.2}$  & 0.155$^{+0.073}_{-0.041}$ & \citet{feng22} \\ 
 \noalign{\smallskip}
 $\nu$ Oct [A]  & $\nu$ Oct [B] & b & 2.6 & 1.25$\pm$0.05 & 0.11$\pm$0.02 & \citet{ramm16,ramm21} \\ 
 \noalign{\smallskip}
 \hline
 \end{tabular}
 }
 \label{tab:systems_with_sa20}
\end{table*}

\subsection{Eccentricities}
\label{sec:eccentricities}

\begin{figure}
 \centering \includegraphics[width=1\linewidth, angle=0]{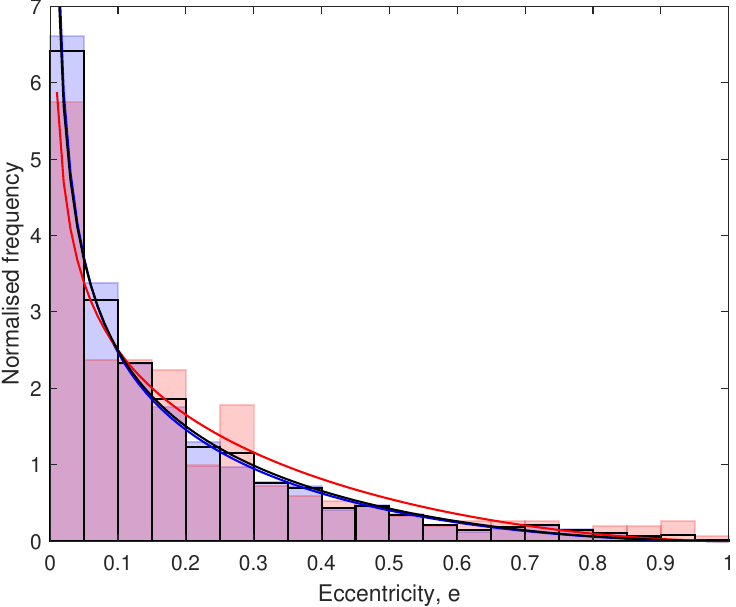}
 \caption{Distributions of eccentricities of planetary orbits and fitted beta distributions.  Black open bins: joint sample (1332 planets in single plus multiple stars); blue filled bins: single-star sample (1029 planets); red filled bins: multiple-star system sample (303 planets). The histograms are normalised to have a unit area. The three curve fittings (black for the joint sample, blue for single-star systems, and red for multiple-star systems) have beta distributions with different $\alpha$ and $\beta$ parameters. See text for a description.}
 \label{fig:ecc_global}
\end{figure}

The previous section prepared the ground for this one where the inter-comparison of the planet eccentricity distributions gets more complicated and uses new tools introduced here.
Except for the unconfirmed astrometric exoplanet candidate around one of the two stars in HD~176051AB and for $\gamma$~Cep~Ab (which was originally discovered by \citealt{campbell88} but confirmed 15 years later by \citealt{hatzes03}), all the planets in Table~\ref{tab:systems_with_sa20} have eccentricities $e$ that are significantly different from null.
Furthermore there are two planets with $e >$ 0.7 and a third planet with $e \approx$ 0.48.
Therefore, we addressed the question of whether the multiplicity of the host stellar system has any effect on the eccentricities of the planetary orbits, as previously suggested by other authors \citep[e.g.][]{eggenberger04b,raghavan06,lester21}. %Instead of comparing the mean values for the $e$ between planets in single and multiple stars, 
In Fig.~\ref{fig:ecc_global} we show the distribution of eccentricities of our combined sample of planets around single and multiple stars. 
Apparently, there are more planets with low eccentricity in single systems and more planets with high eccentricities in multiple systems.

To assess the reliability of this possible effect, we first modelled the observed distribution of planet eccentricities in both single and multiple star samples using a unique probability distribution.
We found that the beta distribution, the simplest distribution representing a continuum variable taking values between 0 and 1, can indeed be used to model such distribution.
The likelihood can, therefore, be written as:

\begin{equation}
P(e|\alpha,\beta)=\frac{e^{\alpha-1}(1-e)^{\beta-1}}{B(\alpha, \beta)},
 \label{eqn:beta}
\end{equation}

\noindent where $B(\alpha, \beta)$ is the beta function as a function of gamma functions:

\begin{equation}
B(\alpha, \beta) = \frac{\Gamma(\alpha) \Gamma(\beta)}{\Gamma(\alpha+\beta)}.
 \label{eqn:gamma}
\end{equation}

In order to fit a beta distribution to these data, we constructed a Bayesian Markov Chain Monte Carlo (MCMC) model written in Stan\footnote{\url{https://mc-stan.org}} \citep{carpenter17}. 
Stan is a programming language that makes use of the Hamiltonian Monte Carlo no U-turn sampler (HMC+NUTS) algorithm \citep[][]{neal11,hoffman11} to perform Bayesian modelling and inference. 
This programming technique allowed us to introduce in the calculations the errors in the measured eccentricity, which are typically asymmetrical.
In some cases, only an upper value for the eccentricity is available. 
This lack was solved by incorporating asymmetric Gaussian distributions in our Stan program. 
For the few planetary orbits without an estimation of the eccentricity error we assumed a typical value of 0.1.

Instead of working with the usual $\alpha$ and $\beta$ parameters of the beta distribution, we parameterised it using the more intuitive parameters mean, $p=\alpha/(\alpha+\beta)$, and concentration, $\theta=\alpha+\beta$. 
First, for the joint sample of 1332 planetary orbits in 902 single and multiple systems (1029 planets around 687 single stars, 303 planets in 215 multiple systems, including the 3 binary systems with detected planets around both stars), we derived the  
posterior probability distributions for the parameters and obtained $p=0.1719\pm0.0050$  and $\theta=3.42\pm0.18$. 
The quoted errors are the standard deviation of these distributions, although they are not necessarily Gaussian. 
We did not fit the curve to the represented binned data, but to the original unbinned eccentricities using their measurement errors.
The resulting beta distribution is plotted in black in Fig.~\ref{fig:ecc_global}. 

To be able to assess the possible effect of the multiplicity of the host system on the eccentricity we fitted two additional models. 
In the second one we fitted the $p$ and $\theta$ parameters separately for planets in single and multiple systems. 
The result was that there is no statistically significant difference in the $\theta$ values between both samples, with a mean difference of $\Delta\theta = |\theta_{\rm single} - \theta_{\rm multiple}| = 0.22\pm 0.21$. 

We finally fitted a third model in which the $\theta$ value was the same for both samples, whereas the $p$ parameter was allowed to vary between them. 
The posterior parameters for this model were $\theta = 3.45\pm0.18$, $p_{\rm single}=0.1627\pm0.0058$, and $p_{\rm multiple}=0.203\pm0.012$. 
Fig.~\ref{fig:ecc_global} also displays the distributions of eccentricities of planets in single and multiple systems and their corresponding beta fits (with $\theta_{\rm single} = \theta_{\rm multiple}$), in blue and red respectively.

\begin{figure}
 \centering \includegraphics[width=1\linewidth, angle=0]{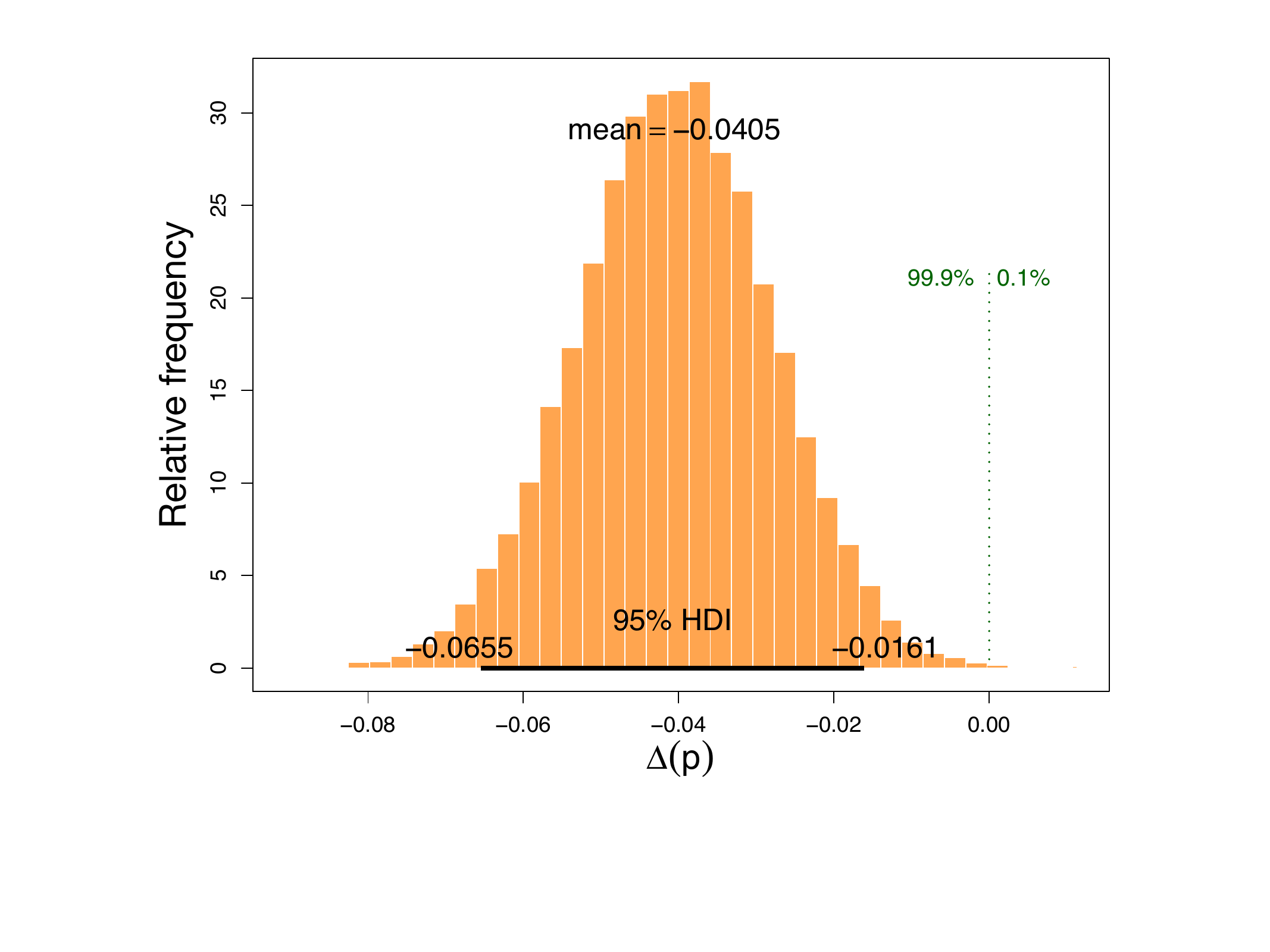}
 \caption{Posterior probability distribution for the difference in the mean $p$ parameter of the beta distributions fitted to the eccentricity distributions of planetary orbits in simple and multiple stars. Here $\Delta p  = p_{\rm single} - p_{\rm multiple}$, and $\theta_{\rm single} = \theta_{\rm multiple}$. The thick horizontal black line represents the 95\% HDI (see text), while the black labels indicate the upper and lower values of the 95\% HDI and the mean $\Delta p$. 
 The vertical green dotted line indicates the location of the null difference, while the percentages show %the magnitudes of 
 the relative areas of the distribution at both sides of this line. 
 This figure was created using a modified version of the software provided by \citet{kruschke15}.}
 \label{fig:ecc_deltap}
\end{figure}

To carry out an objective comparison among the three models, we applied an information criterion, namely the Watanabe-Akaike Information Criterion \citep[WAIC;][]{watanabe13}. 
The particular values obtained for the first (same $p$ and $\theta$ parameters), second (different $p$ and $\theta$), and third (fixed $\theta$, different $p$) were $-2891.8$, $-2908.2$, and $-2909.3$, respectively.
We obtained identical results with the Leave-One-Out Cross-Validation criterion \citep[LOO-CV;][]{vehtari17}.
Therefore, the third model, with shared $\theta$ and different $p$ parameters, was indeed the best model representing the data according to this criterion. 
The analysis with the MCMC technique provides an estimate of the unknown parameters of the fitted model for each step of the Markov chains. In this case we run four separate chains with 12\,500 points each and, therefore, obtained 50\,000 estimates for the difference $\Delta p=p_{\rm single} - p_{\rm multiple}$. 
A histogram of these values is represented in Fig~\ref{fig:ecc_deltap}. 
According to \citet{gelman14}, who thoroughly discussed the Bayesian data analysis methods, this distribution is an unbiased true representation of the actual posterior distribution of the parameters. 
The thick horizontal line in the figure shows a 95\% highest density interval (HDI). The HDI is the Bayesian counterpart of the classical confidence interval, and it is calculated as the narrowest interval that contains a certain percentage of the probability density. In other words, it encompasses the most credible values of the distribution, being not necessarily symmetrical nor centred on the arithmetic mean (see, for example, \citealt{kruschke15} for a definition and discussion of the HDI).

In our case we obtained a posterior distribution that implies $\Delta p = -0.041\pm0.013$. 
In Fig.~\ref{fig:ecc_deltap} we compare the extent of the 95\% HDI bar with the position of the zero value.  
Only 0.1\% of the posterior estimates are positive, and, therefore, our conclusion is that $\Delta p < 0$ with a significance level of 0.001 in conventional statistics.  
Thus, we concluded that planetary orbits in multiple stellar systems do exhibit significantly larger eccentricities than those around single stars.
This result is expanded, nevertheless, in the next section.

\subsection{Semi-major axes, separations, and eccentricities}
\label{sec:semi-major_sep_eccen}

Given the result obtained in Sect.~\ref{sec:eccentricities}, one would expect that the physical separation between the stars in the closest multiple systems may have an effect on the eccentricities of planetary orbits, being this effect negligible for the widest pairs. 
To check and quantify this hypothesis, we fitted a model in which we allowed the mean $p$ of the beta distribution of the eccentricities $e$ to vary with the physical separation $s$ between stars.
For this purpose we constructed a generalised lineal model. 
This kind of models is useful to explore regressions with dependent variables not following a Gaussian distribution, such as the orbit eccentricities in our case, although its use in astrophysical research is limited.
We refer the reader to \citet{desouza15a,desouza15b}, \citet{elliot15}, and \citet{hilbe17} for discussion of this statistical technique in the context of astrophysical problems, including some useful examples.

In our case we fitted a generalised model in which the possible effect of physical separation in logarithmic units, $\log{s}$, was introduced by computing a linear predictor $\eta$:

\begin{equation}
\eta=\beta_1 + \beta_2\log{s},
\label{eqn:eta}
\end{equation}

\noindent which was related to the mean $p$ of the beta distribution by a logit link function \citep{berkson44,mcfadden73}:

\begin{equation}
\eta \equiv {\rm logit}~p=\log\frac{p}{1-p}
\label{eqn:logit}
\end{equation}

\begin{figure*}
 \centering \includegraphics[width=1\linewidth, angle=0]{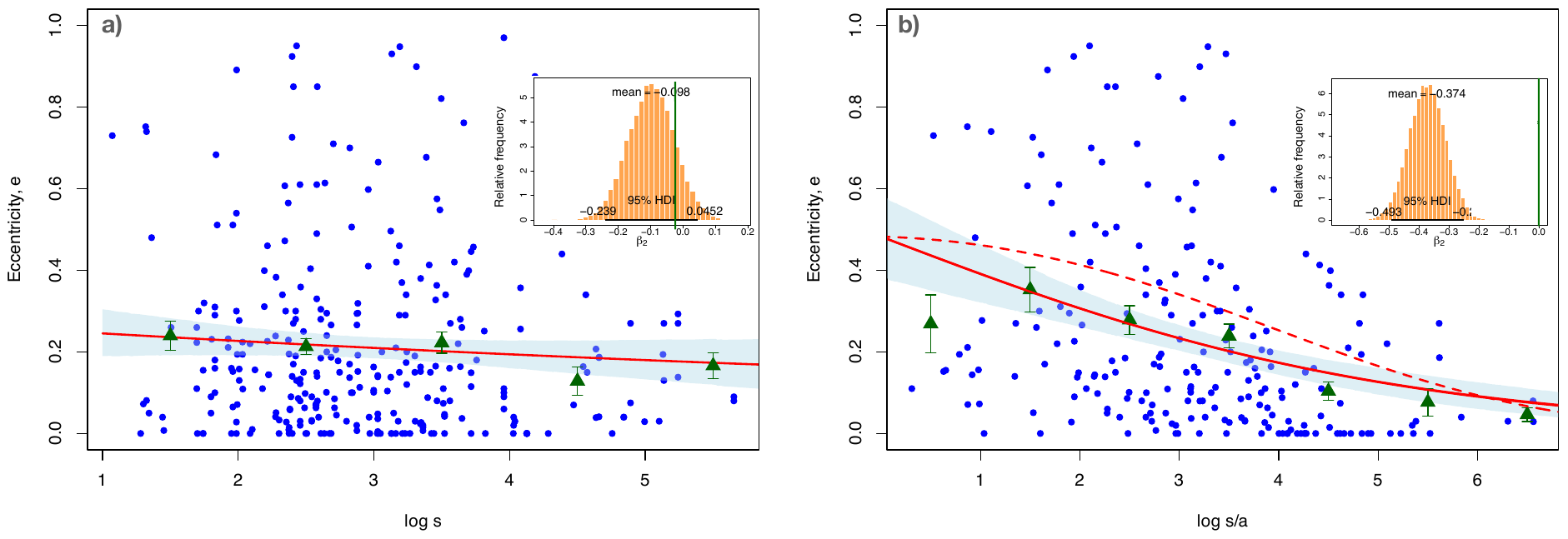}
 \caption{Planet eccentricity as a function of star-star separation (panel a, \textit{left}, in au) and of ratio between star-star separation and semi-major axis (panel b, \textit{right}).
 The red lines show the predicted mean value $p$ of the beta distributions as a function of the dependent variable, in logarithmic scale. 
 The blue shades around these lines correspond to their 95\% HDI. 
 The insets show the corresponding posterior probability distribution of the $\beta_2$ parameters, and the vertical green lines mark $\beta_2=0$. 
 Green triangles mark the mean eccentricities in bins of 1\,dex, together with error bars for their formal errors. The dashed red line in panel b shows the result of a generalised model using a quadratic fit.}
 \label{fig:panelsep}
\end{figure*}

We introduced the link function to transform from the unbounded scale of $\eta$ to the bounded range of $p$ between 0 and 1. 
The $\theta$ parameter of the beta distribution was not allowed to vary with the stellar separations, though. 
To perform the fit we wrote another MCMC model in Stan to derive posterior probability distributions for the three parameters $\theta$, $\beta_1$, and $\beta_2$. 
These distributions can be summarised as $\theta=1.981 \pm 0.160$,  $\beta_1=-1.034 \pm 0.226$, and $\beta_2=-0.098 \pm 0.072$. It is already apparent from the estimate for $\beta_2$ that the effect of the stellar separation on the eccentricities is not statistically significant. This is more clearly seen in the left panel of Fig.~\ref{fig:panelsep}, where we represent the prediction for the $p$ parameter, and its 95\% HDI in the $e$--$\log{s}$ plane. 
To guide the eye, and although they were not used in the fitting procedure, we included in Fig.~\ref{fig:panelsep} the mean eccentricities in bins of 1\,dex.

The above result is not surprising since one would expect that the effect of star separation $s$ on the orbit eccentricities should not depend just on the absolute value of $s$, but on its relative value compared to the star-planet separation. 
To confirm this hypothesis, we repeated the above analysis but replaced $s$ by the ratio $s/a$ between the separation between stars and the semi-major axis of the planetary orbit. 
As before, in the case of multiple planetary systems we used the $a$ value of the outermost planet, and in the case of triple and quadruple systems we used the $s$ value of the innermost star (or white dwarf or brown dwarf). 
The results are illustrated by the right panel of Fig.~\ref{fig:panelsep} and summarised by the following parameters: $\theta=2.09 \pm 0.20$,  $\beta_1=-0.07 \pm 0.21$, and $\beta_2=-0.374 \pm 0.062$. 
In the $e$--$\log{s/a}$ plane, the $\beta_2$ parameter is significantly different from zero; the equivalent significance level in conventional statistics is $<2\times10^{-5}$ (i.e. beyond a 4$\sigma$ effect). 
As a result, for a fixed semi-major axis $a$, planets in multiple systems with shorter star-star separations $s$ tend to exhibit larger eccentricities $e$.

The mean binned eccentricities plotted in the right panel of Fig.~\ref{fig:panelsep} suggests a flattening of the relation for the lowest values of $\log(s/a)$. 
To assess the reliability of this effect we added a quadratic term to the relation between $\eta$ and $\log(s/a)$, thus fitting the following relation:

\begin{equation}
\eta = \beta_1 + \beta_2 \log(s/a) + \beta_3  \log^2(s/a)    
\end{equation}

\noindent The result, which is also shown in the right panel of Fig.~\ref{fig:panelsep}, confirms the suspected flattening. However, according to the values of the WAIC information criterion, this expanded model does not improve the linear one, which we kept afterwards.

\subsection{Number of planets in single and multiple systems}
\label{sec:planet_frequency}

We also investigated the planet rate in single (with one star) and multiple systems (with two, three, or four stars).
The arithmetic mean numbers of planets around the 687 single and 215 multiple stars in our two samples are 1.51 and 1.41, respectively, which suggest that single stars tend to host slightly more planets. 
To test whether this measured difference is statistically significant, we applied the MCMC technique, programmed in Stan, to fit Poisson distributions to the detected number of planets in each sample. 
Since the two samples do not include cases with no planets orbiting the host star, we used a zero-truncated Poisson (ZTP) distribution \citep{hilbe17}, which corrects the classical Poisson model to exclude the possibility of observations with zero counts. In particular, 
if \textit{N} denotes the random variable representing the number of planets, and $\lambda$ is the parameter of the Poisson distribution, the ZTP probability distribution function takes the form:

\begin{equation}
P(N=k)=\frac{\lambda^k e^{-\lambda}}{k!\,(1-e^{-\lambda})},
\label{eqn:ztp}
\end{equation}

\noindent where the corrected mean number of planets $\mu$ is:
\begin{equation}
\mu = \frac{\lambda}{1-e^{-\lambda}}.
\label{eqn:mu_ztp}
\end{equation}

\noindent Here $e^{-\lambda}$ is the expected number of zeroes in a Poisson distribution with parameter $\lambda$.

\begin{figure}
 \centering \includegraphics[width=1\linewidth, angle=0]{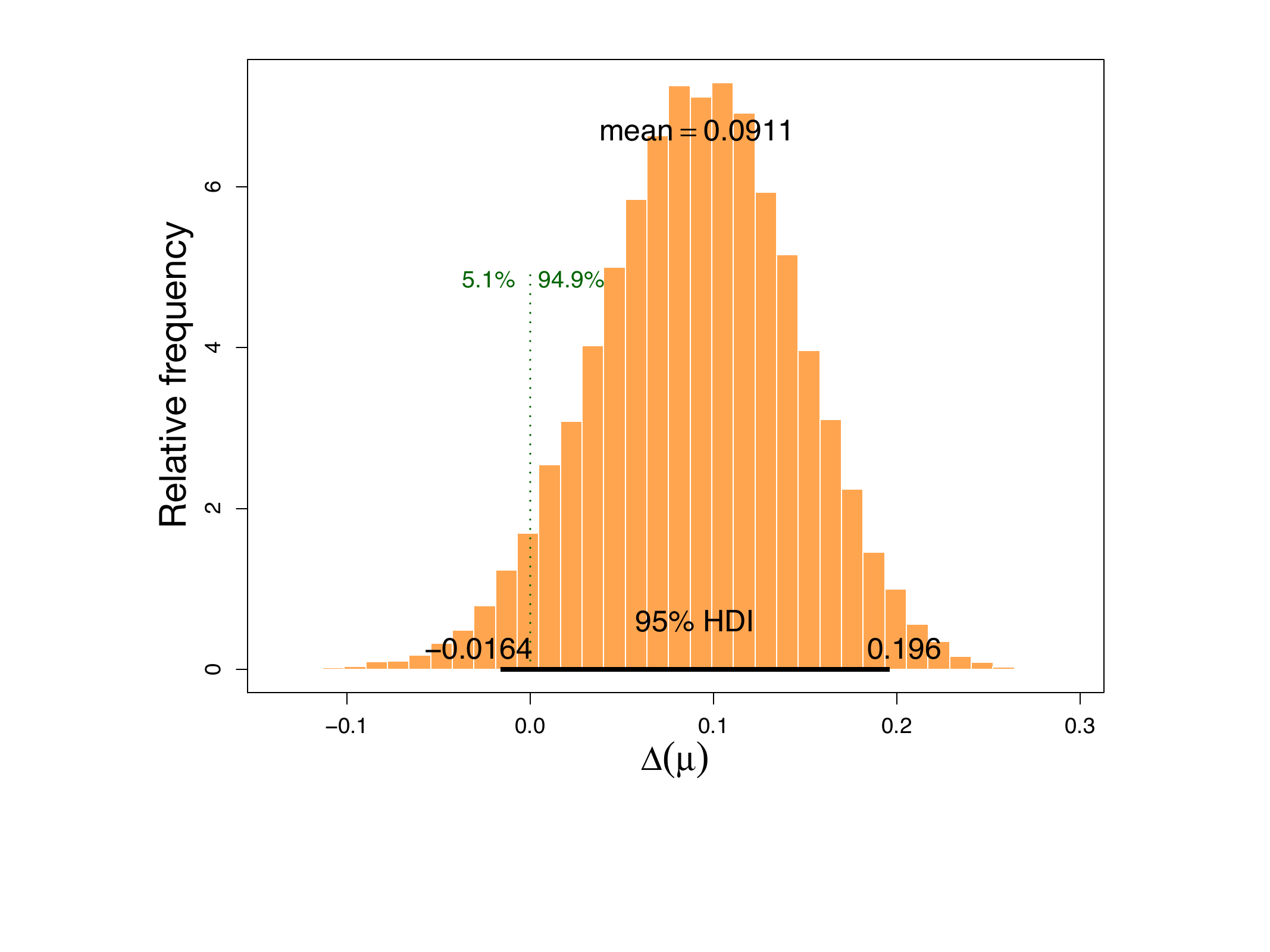}
 \caption{Same as Fig.~\ref{fig:ecc_deltap} but for the difference in the number of planets in single and multiple systems, $\Delta \mu = \mu_{\rm single} - \mu_{\rm multiple}$.}
 \label{fig:difmeanp}
\end{figure}

The corresponding means of planets per host star are $\mu_{\rm single}$ = 1.508 $\pm$ 0.029 and $\mu_{\rm multiple}$ = 1.416 $\pm$ 0.046, virtually identical to the arithmetic means. 
The posterior probability distribution of the difference of the two means is represented in Fig.~\ref{fig:difmeanp}, which displays a mean offset of $\Delta\mu=0.091\pm0.054$. 
This difference is only marginally significant and hovers at the edge of the 95\% confidence interval. 
As a conclusion, although the data suggest a larger number of planets around single stars when compared to multiple systems, the statistical significance of this effect is weak and we could not derive a firm conclusion.

One caveat of the previous analysis is that the applied ZTP distribution is not completely appropriate since, as it is often the case in real observational data, the observed dispersion in $N$ is larger than the one expected from a Poisson distribution (i.e. $\sigma>\sqrt{\lambda}$). 
To solve this problem, we used an alternative probability distribution, namely the negative binomial, which is similar to the Poisson distribution but with one more parameter to account for a possible overdispersion. 
We again refer the reader to \citet{desouza15b} and \citet{hilbe17} for descriptions of the use of the negative binomial distribution in astrophysical contexts. 
To analyse the possible effects of this more reliable distribution, we constructed an MCMC model introducing a zero-truncated negative binomial distribution and applied it to our data. 
The result is that the mean offset in the number of planets between both samples becomes $\Delta\mu=0.088\pm0.066$. 
This difference is even less significantly different from zero than our estimate with the Poisson distribution and, therefore, we reinforce our previous conclusion of a non-significant effect of the stellar multiplicity on the number of planets per star.

Given the results of Sect.~\ref{sec:semi-major_sep_eccen}, in which we showed how the increase of eccentricities is more apparent in low-$s/a$ systems (i.e., systems in which the star-star separation is not much wider than the size of the planetary orbit), a possible effect of a lower number of planets for multiple systems could be hindered by the inclusion of high-$s/a$ systems, which could be more similar to single systems. %Following a referee's suggestion, to 
To check this hypothesis, we constructed a generalised linear model in which the mean parameter of the Poisson distribution was allowed to vary with $\log(s/a)$. 
As in Sect.~\ref{sec:separations}, in the case of multi-planet systems, we used the $a$ value of the outermost planet and $s$ of the closest stellar companion, so $s/a$ should be read as ${\rm min}(s/a)$.
In particular, we used an exponential link function to transform from the unbounded scale of a linear relation to the positive scale of the parameter of the Poisson distribution, i.e.:

\begin{equation}
   \lambda = e^{\beta_1 + \beta_2\log(s/a)} 
\end{equation}

The result for the slope parameter of the linear relation is $\beta_2=-0.015\pm0.080$. Therefore, it is not significantly different from zero, and we concluded that there is no apparent effect of $s/a$ on the observed number of planets.
A computation of the corresponding WAICs also indicates that this model is worst than the constant $\lambda$ model.

\subsection{Planet masses in single and multiple systems}
\label{planet_masses}

\begin{figure}
 \centering \includegraphics[width=1\linewidth, angle=0]{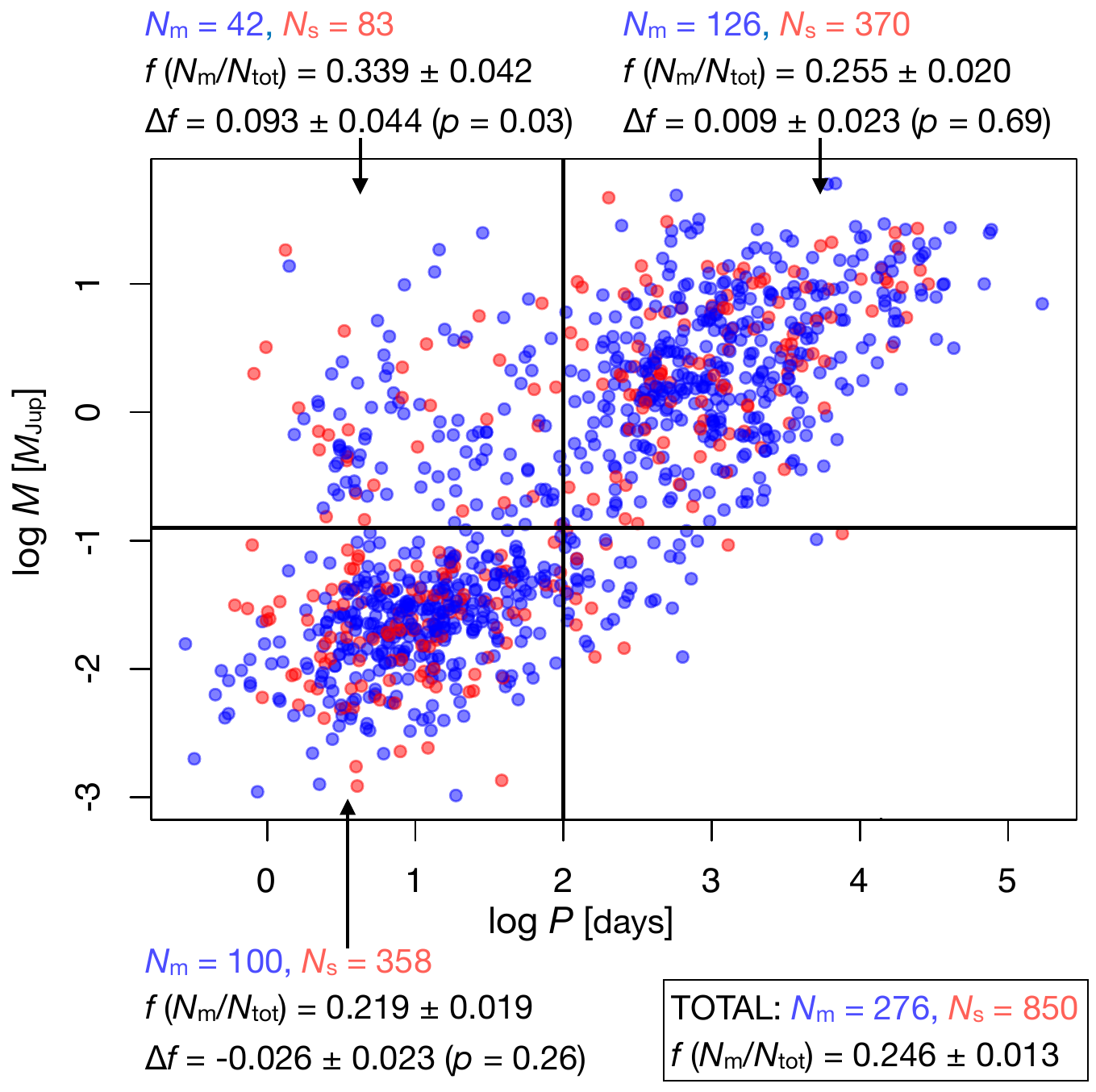}
 \caption{Period-mass diagram for planets in single (red circles) and multiple systems (blue circles). 
 The horizontal black line divides the sample into high- and low-mass regimes using a cutoff of 0.1258\,M$_{\rm Jup}$ (40\,M$_\oplus$), while the vertical black line at $P$\,=\,100\,d marks the position of the arithmetic mean of the distribution in orbital periods. 
 The information showed for each quadrant is the number of planets in multiple, $N_{\rm m}$, and single systems, $N_{\rm s}$, the fraction of planets in multiple systems with respect to the whole sample, $f=N_{\rm m}/(N_{\rm m}+N_{\rm s})$, the difference $\Delta f$ between this ratio and the value computed for the total sample ($f_{\rm tot}=0.246$), and, in parenthesis, the statistical significance $p$ of the differences. 
 No information is offered for the bottom-right quadrant due to the insufficient number of planets. 
 }
 \label{fig:period_masses}
\end{figure}

In this section, we explore whether stellar multiplicity has an impact on the locus of planets in the orbital period-planetary mass diagram.
For example, \citet{fontanive19} and \citet{fontanive21}, while not offering specific statistics, suggested that the separation of high-mass planets is influenced by stellar multiplicity. 
With the expanded and enhanced sample presented in our work, we can address these concerns with greater confidence.
In Fig.~\ref{fig:period_masses} we display the period--mass diagram where we discriminate between planets in single and multiple systems. 
As shown in the plot, our sample comprises 850 planets around single stars and 276 in multiple systems with all data, which account for a 25\% of the total sample. 
At a first glance, Fig.~\ref{fig:period_masses} suggests that high mass planets with short periods may be relatively more frequent in multiple systems. 
To verify this hypothesis, we divided the plane into four quadrants, as depicted in Fig.~\ref{fig:period_masses}, and determined a number of parameters, together with their errors, by fitting binomial distributions using the MCMC technique. 
We found a marginally significant difference only in the upper-left quadrant, which suggested that high-mass planets ($M > 40\,{\rm M}_\oplus$) in close orbits ($P < 100$\,d) appear to be relatively more frequent in multiple systems than around single stars.
Other authors, such as \citet{eggenberger04} and \citet{fontanive21}, have also concluded that the most massive short-period planets (with masses greater than 2\,M$_{\rm Jup}$) also tend to orbit in multiple star systems.

\begin{figure}
 \centering \includegraphics[width=1\linewidth, angle=0]{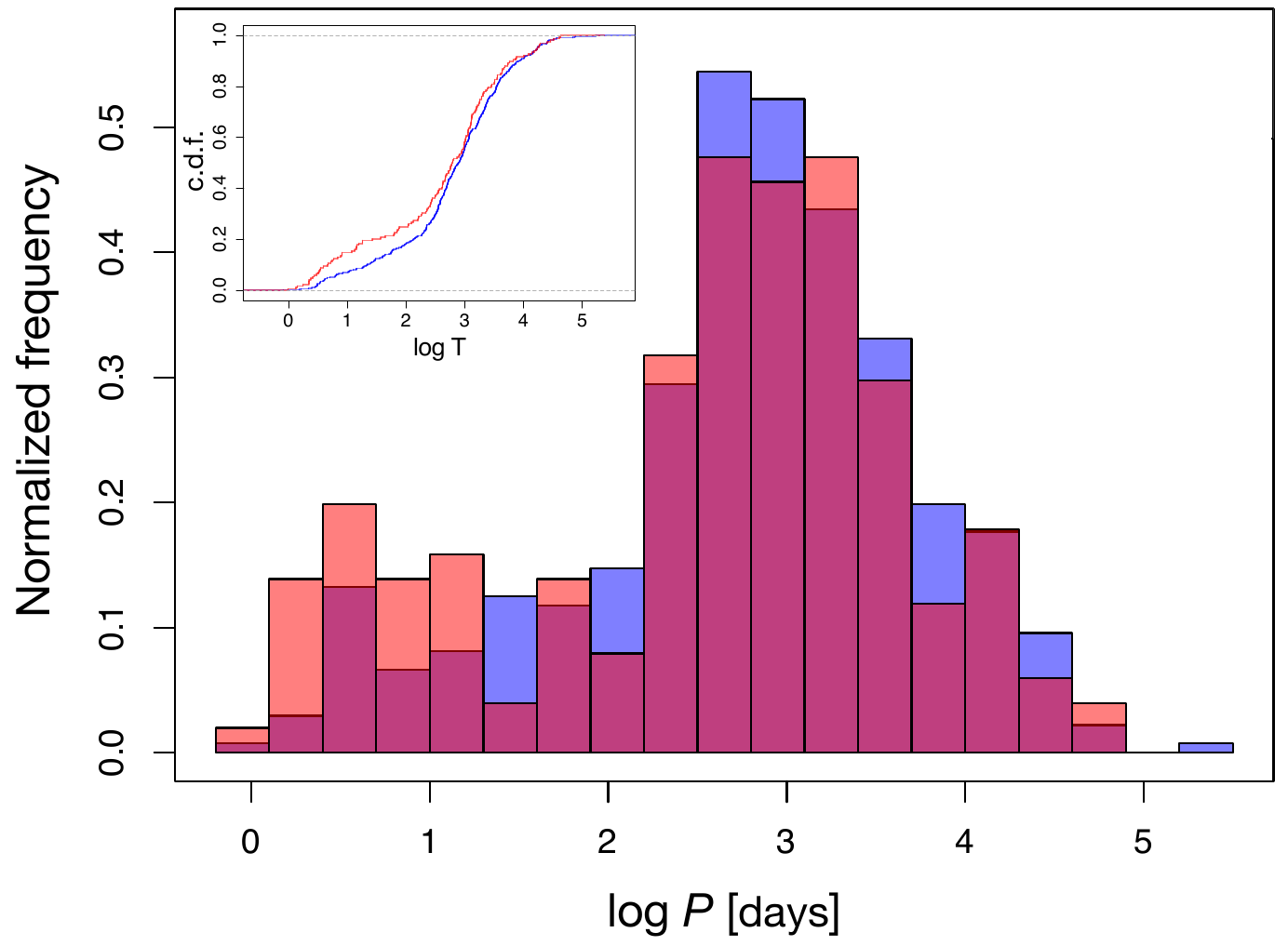}
 \caption{Histogram of the distributions of orbital periods for high-mass ($M>40$\,M$_\oplus$) planets in single (blue bars) and multiple (orange bars) systems. The inset of the figure compares the cumulative distribution functions of both subsamples.}
 \label{fig:hist_period}
\end{figure}

To delve deeper into these potential findings, in Fig.~\ref{fig:hist_period} we present histograms illustrating the distribution of orbital periods for high-mass ($M$\,>\,40\,M$_\oplus$) planets in both single and multiple systems. 
The relative increase in the proportion of planets with low orbital periods in the latter systems becomes apparent, as suggested by the previous quadrant analysis.
To assess the reliability of this observation, we compared the cumulative distribution functions of both subsamples, as depicted in the inset of Fig.~\ref{fig:hist_period}. For this purpose, we conducted an Anderson-Darling test \citep{anderson52}, which is a modification of the Kolmogorov-Smirnov test, but more suitable for situations where greater emphasis must be placed on the tails of the distribution\footnote{\url{https://asaip.psu.edu/Articles/beware-the-kolmogorov-smirnov-test}}, as it is the case with our data.
The outcome of this test indicates that the statistical significance of the difference between both distributions is $p=0.054$. Consequently, we cannot draw a definitive conclusion, and larger samples are evidently required to evaluate the reliability of this trend.

\section{Discussion}
\label{sec:discussion}

\subsection{Missing multiple systems}
\label{sec:missing_multiple_systems}

We evaluated the existence of unidentified multiple systems in our sample.
First, our common proper motion and parallax search was limited by the \textit{Gaia} completeness at $G \sim$ 20.3\,mag.
According to Table~7 of \citet{smart19}, our \textit{Gaia} search was complete down to spectral type M8$\pm$1\,V up to 100\,pc, and L8$\pm$1 up to 10\,pc.
As a result, we were expected not to be able to detect ultracool dwarfs M9\,V and later in the whole surveyed volume, nor T- and Y-type brown dwarfs just beyond 10\,pc.
Actually, we identified an L1 ultracool dwarf as a wide companion to HD~1466 in Tucana-Horologium, but it is overluminous because of its youth (Sect.~\ref{sec:remarkable_systems}).
While we might have conservatively stated that our \textit{Gaia} search for resolved companions was virtually complete in the stellar domain down to about 0.1\,M$_\odot$ \citep{cifuentes20}, it was far from being complete for the least massive stars and the whole substellar domain.
Furthermore, most of the brown dwarf companions analysed here came from WDS or literature works that employed deep adaptive optics imaging (e.g. GJ~229\,B; \citealt{nakajima95}).
Additional wide stellar companions with spectral types earlier than M8\,V might have also been uncatalogued by \textit{Gaia} in crowded regions at low galactic latitudes \citep{reyle21}.

Second, there may be more unidentified multiple systems, although hidden in plain sight.
Exoplanet-host stars have naturally been monitored and validated with high-resolution spectroscopy in search for close unresolved binaries.
Except for a few cases with long-term radial-velocity trends superimposed on short-term exoplanetary signals (e.g. 51~Peg itself), most host stars are known to have only exoplanetary companions at close separations.
However, their resolved companions, usually fainter and with later spectral types, have in general not been analysed so thoroughly, so it may happen that some of the identified double systems actually are triple systems (and triple systems are quadruple systems, and so on).
Furthermore, the proverb ``these things always come in threes'' seems to apply well to multiple stellar systems through a major prevalence of wide triples over wide binaries \citep{basri06,tokovinin06,caballero07,cifuentes21}.

There have been only a few papers with exhaustive searches for companions of very different masses and at very different projected separations from the primary stars.
For example, \citet{caballero22} compiled CHARA interferometry, NICMOS/\textit{Hubble} Space Telescope and PUEO/Canada-France-Hawai'i Telescope high-resolution imaging, CARMENES/3.5\,m Calar Alto and MAROON-X/Gemini high-resolution spectroscopy, and \textit{Gaia} spectrophotometry, and ruled out the presence of stellar and high-mass brown-dwarf companions to the exoplanet-host star Gliese~486 from the limit of the \textit{Hubble} observations at 24--32\,au up to 100\,000\,au and of planets with minimum masses of $\sim$30\,M$_\oplus$ with periods up to 20\,000 days ($\sim$11\,au).
Other planetary systems have not been the subject of such a massive observational effort, but reasonable upper limits to the presence of additional companions exist for many of them.
As a result, some of the 687 single stars with planets may not be single, but there are high chances that the great majority of them, especially those at the closest distances, do not have stellar companions at close or wide separations.

\subsection{Eccentricity: nature or nurture}
\label{sec:eccentricity}

One of the main results of our work is that the orbital eccentricity, $e$, of known planets in multiple stellar systems does not depend on the projected physical separation between component stars, $s$, but on the ratio between this separation and the planet semi-major axis, $a$.
Therefore, the smaller the $s/a$ ratio, the more probable the planet $e$ is significantly larger than zero.
However, in spite of this significant result after so many previous searches and analyses, we are still far from understanding the nature-nurture dilemma of eccentric planets in close multiple stellar systems (Sect.~\ref{sec:introduction}) or the role of the modulation of the orbital eccentricity, which strongly depends on the relative inclination between the plane of motion of the planet and that of the binary \citep{mazeh97}.
This relative inclination is known for very few systems, if any, as the planet must (in general) transit and the stellar binary orbit must have an astrometric solution.
The collection of multiple stellar systems with exoplanets in Table~\ref{tab:systems1}, especially those with short projected angular separations, is an excellent input for forthcoming determinations of astrometric orbit parameters with periods of a few years.
Such determinations will open the door to further dynamical analyses of systems with stellar and planetary companions orbiting primary stars at different angles, from which to extract conclusions on their formation, evolution, and stability.

We also made a visual inspection of planet $e$ as a function of total mass of the system, and did not find any hint of relation that could be statistically analysed.
It happened the same when we inspected the exoplanets orbital periods and the number of planets in system as a function of total mass.
However, trying to analyse the detectability of exoplanets and, therefore, the number and (minimum) mass of exoplanets per system as a function of parameters of only the planet-host stars, such as their mass, spectral type, and degree of activity, is out of the scope of this work \citep[e.g.][]{sabotta21}.

\subsection{Multiplicity function and future exoplanets surveys}
\label{sec:multiplicity_function_future_exop_surveys}

We went on by comparing our results with those from the literature.
The stellar MF of exoplanet systems (i.e. the fraction of exoplanet systems with stellar companions) has been determined since the very beginning of exoplanetology at about 20\% \citep{bakos06,bonavita07,mugrauer07}. 
More precise determinations have varied around the narrow interval of 23\% (30 multiple systems in 131 planetary systems; \citealt{raghavan06}) and 23.2 $\pm$ 1.6\% (218 multiple systems in 938 planetary systems; \citealt{fontanive21}), compatible with the value of at least 12\% proposed by \citet{roell12}.
In our case, after discarding pairs in open clusters and associations and with ultracool dwarf companions, we determined an MF of exoplanet systems of 21.6 $\pm$ 2.9\% (212 multiple systems in 899 planetary systems).
We stress on three issues in the comparison with the recent value of \citet{fontanive21}: 
($i$) the actual MF of exoplanet systems must be slightly larger because all searches, including ours, are incomplete at the latest spectral types (Sect.~\ref{sec:missing_multiple_systems});
($ii$) the identical value but the larger uncertainty of our measurement in spite of the similar numbers.
In particular, we used the Wald interval \citep{agresti98} assuming Poisson statistics for computing our uncertainties with a Wald 95\% confidence interval (Wald interval\,$=(\lambda - 1.96\sqrt{\lambda/n}, \lambda + 1.96\sqrt{\lambda/n})$, where $\lambda$ is the number of successes in $n$ trials);
and ($iii$) they explored a volume eight times larger than us, since they studied systems up to 200\,pc, and we did it up to 100\,pc.
Our value of 21.6 $\pm$ 2.9\% must necessarily be compared with MFs of nearby stars regardless planet presence.
Since the MF depends on spectral type, the comparison should be done with that of field stars of spectral types between late F and mid M, as most radial-velocity and transit surveys during the last three decades have focused on them.
All reference MF values enumerated in Sect.~\ref{sec:introduction} are larger than our MF.
Although there have been claims that the influence of a binary companion disallows from hosting planetary systems, either because of formation or evolution \citep[e.g.][]{wang14a,kraus16}, we instead ascribe our lower multiplicity fraction to the aforementioned exoplanet survey bias (\citealt{winn15} and the CARMENES examples; Sect.~\ref{sec:results_planetary_systems}).

There is no exoplanet survey free of multiplicity bias, unless especially designed to overcome it.
For example, one could envisage a search for smaller-amplitude signals in radial-velocity and light curves of known spectroscopic and eclipsing binaries, which are generally discarded in the first stages of exoplanet surveys \citep{garciamelendo11,gillon17,jeffers18,tal-or18,parviainen19,kristiansen22}.
Although several major transiting surveys are or will be greatly affected by multiplicity because of their poor spatial resolution, such as TESS \citep{ricker15} and PLATO \citep{rauer24},
wide multiplicity barely affects radial-velocity exoplanet searches.
Actually, some of the stars in wide multiple systems monitored by CARMENES, with companions at more than 5\,arcsec, eventually turned out to host new exoplanets \citep[e.g.][]{trifonov18,reiners18a,kaminski18,gonzalezalvarez20,kossakowski21}.
The CARMENES limit at 5\,arcsec was defined from the size of the spectrograph's optical fibre projected on the sky, of 1.5\,arcsec, the typical seeing at the Calar Alto observatory, and the maximum allowed spectral contamination by a close companion \citep{quirrenbach14,cortes17b}.
Similar minimum separations have been defined for other radial velocity searches with HARPS \citep{bonfils13} or SPIRou \citep{moutou17b,fouque18}, just too put two comparable examples.
A separation of 5\,arcsec is, however, too short for TESS and PLATO, which have pixel sizes of 15.6--21.0\,arcsec and, therefore, they need compulsory \textit{Gaia}, lucky imaging, or adaptive optics data for validating transiting targets or even pre-launch scientific prioritising targets \citep{collins18,clark22,nascimbeni22}.
As a result, the crux of the matter is perhaps the definition of a boundary between ``wide'' and ``close'' multiples, which depends on the exoplanet survey.
For example, one of the key exoplanet surveys of the next decades will be the Habitable Worlds Observatory\footnote{\url{https://habitableworldsobservatory.org}} \citep[HWO;][]{clery23},
for which preliminary target lists are starting been defined right now \citep{tuchow24,harada24}.
Although the HWO scientific requirements and, therefore, mission specifications are far from being fixed, a wide multiple stellar system (i.e. one in which a stellar companion does not affect observations of host stars) must necessarily be less than 1\,arcsec for HWO.
The same shall happen to the Large Interferometer For Exoplanets\footnote{\url{https://life-space-mission.com}} \citep[LIFE;][]{quanz22}.
The bulk of the resolved stellar systems have angular separations greater than that value (there is a recent \"Opik diagram in Fig.~1 of \citealt{gonzalezpayo23}).
However, the astronomical community must make a joint effort to complement forthcoming \textit{Gaia} DR4 data for analysing in great detail the innermost arcsecond of the nearest stars regardless their age, either with high-resolution spectroscopy, in search for new spectroscopic multiples, and deep imaging (e.g. \citealt{oppenheimer01,mccarthy04,dieterich12,gauza21}).

\subsection{Spectral types and hierarchy}

\begin{figure}
 \centering \includegraphics[width=1\linewidth, angle=0]{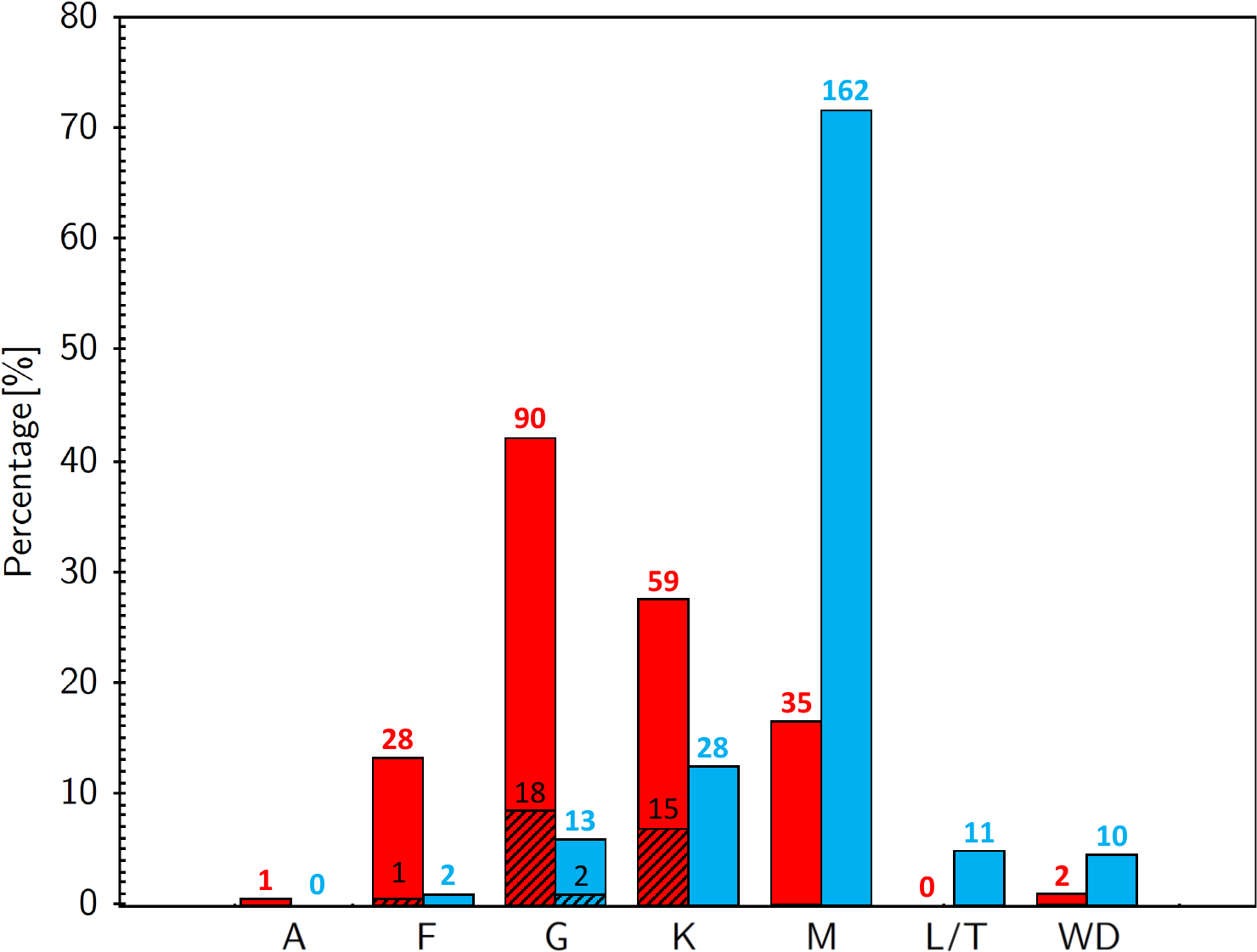}
 \caption{Relative distribution of host stars (red) and companion stars (blue) per spectral type.
 Each bar is labelled with the absolute number of stars per bin.
 The black stripped area represents the number of subgiant and giant stars in each spectral type.
 Compare with Fig.~3 of \citet{fontanive21}.}
 \label{fig:spectral_types}
\end{figure}

Following the comparison with previous works, we also studied the relative distribution of spectral types of all the stars considered in our sample, distinguishing between host and companion stars, as illustrated by Fig.~\ref{fig:spectral_types}. 
Our sample contains stars (and brown dwarfs) from types A, F, G, M, L, T, and white dwarfs (WD). 
As already noticed by \citet{fontanive21}, most host stars are obviously G- and K-type stars, while most companions are M-type stars (see also \citealt{mugrauer07}).
However, we found a larger proportion of late spectral types  as companions, with higher proportion of M stars (71.7\%), and of L and T ultracool dwarfs (4.9\%). 
Many of them were missing in \citet{fontanive21}'s work.
Here the effect of the different maximum survey distances and Malmquist bias played a role. 

Finally, we devote the last part of our discussion to systems hierarchy.
As already mentioned, we found 40 triple and 3 quadruple stellar systems with exoplanets. 
Accounting for the 172 binaries with exoplanets, that makes 18.9\% and 1.4\% of the multiple systems are triple and quadruple.
The values reasonably match those reported by \citet{duchene13} and \citet{reyle21} for field stars in general, regardless they have planets or not.
For example, \citet{reyle21} found that $\sim$20\,\% and $\sim$5\,\% of the multiple systems in the 10\,pc neighbourhood are triple and quadruple+quintuple, respectively. % 69+19+3+2=93
In the case of quadruple systems, \citet{cuntz22} concluded that they are in general made of two groups of binary stars.
This is indeed the case for two of our three quadruples, namely 30\,Ari and HD~18599, but not for the third one centred on HD~1466 (Table~\ref{tab:new_detected_systems}).
This is however a young system in Tucana-Horologium \citep{riedel17,gagne18a} that will likely be disrupted in a few tens of millions of years \citep{caballero09}.
In the case of the triple systems, the 40 of them are made of a close pair plus a separated companion (i.e. AB--C or A--BC), in line with what was already presented by \citet{mugrauer07b}, but with a much poorer statistics of just five triple systems.
As in the case of \citet{cuntz22}, we did not find any system of higher order (i.e. quintuple or higher).

\section{Summary}
\label{sec:summary}

In this study, we compiled all stars with reported planets from the Extrasolar Planets Encyclopaedia and the NASA Exoplanet Archive at less than 100\,pc.
Employing primarily \textit{Gaia} DR3 data supplemented by searches in the Washington Double Star catalogue and available literature,
we identified 215 exoplanet host stars in 212 multiple-star systems with common parallax and proper motion.
During the analysis, we rejected stellar systems that are part of open clusters (Hyades) or OB associations (Lower Centaurus Crux), and planetary systems that have controversial planet candidates (e.g. with minimum masses above the hydrogen burning limit). 
Of the 212 multiple-star systems, 173 are binary (including 2 systems with circumbinary planets), 39 are triple, and 3 are quadruple, as expected from the rate of triple or higher order multiples of field systems.
There are three binary systems with detected planets around both stars.

We identified 17 new companions in 15 systems (10 binaries, 4 triples, 1 quadruple) with known planet host stars.
All the new physically-bound companions have spectral types, either determined by other teams or estimated by us, between K4\,V and L1, except for a white dwarf candidate companion.
Some of the components in the widest systems have been reported to belong to very young stellar kinematic groups, such as Tucana-Horologium, which may indicate that they are currently in the process of disruption by the galactic gravitational potential.

We were able to measure projected physical separations between stars with \textit{Gaia} data in most cases, and to compile key parameters for 276 planets in multiple systems, with which we statistically analysed a series of hypothesis.
First, we computed the ratio between separations and semi-major axes for all multiple systems, and identified the 9 of them with the smallest ratio, $s/a < 20$.
Only one controversial planet candidate, namely $\nu$~Oct~b, poses a challenge to current planet formation and evolution models. 
Next, we constructed a single planet-host star sample for comparison purposes and found a significant difference between the eccentricities of planets in multiple systems and of planets around single stars.
In particular, planets in multiples systems with small $s/a$ ratios have eccentricities that are significantly larger than those of planets with larger $s/a$ ratios.
This result is in line with a number of theoretical predictions; our sample is more complete and our statistics tools are more advanced than most previous observational studies.
Actually, we have a higher proportion of faint M, L, and T companions than previous searches, which provides more statistical robustness to our findings.

Concerning the comparison in the number of planets around multiple and single systems, we  found a possible trend in the sense that the former tend to host a lesser number of planets, although the significance level of this result (around 0.05) precludes us to reach a firm conclusion. 
Also, we detected a greater frequency of high-mass planets in close orbits in multiple systems compared to single ones but, again, this result in on the verge of being statistically significant.
%Besides, we also investigated the number and masses of planets in single and multiple systems, and found no significant differences.
Finally, we estimated a multiplicity fraction of stars with planets of 21.6 $\pm$ 2.9\%, a value close to those reported by similar studies, but also smaller than the multiplicity fraction of field stars regardless they have planets. 
We ascribed this difference not to a different formation or evolution mechanism, but to a well-known observational bias in exoplanet searches, especially in radial-velocity surveys.

One may speculate to extend the survey limit up to 200\,pc or beyond as a future work.
Nonetheless, we must acknowledge the considerable time invested in the individualised analysis of thousands of planetary system candidates, some of which we had to discard.
We rather propose prioritising the continued investigation of stellar multiplicity with planets with a previous work aimed at confirming and characterising a myriad of controversial planetary systems that have polluted past and current analyses. Eventually, we can leverage \textit{Gaia} DR4 for this purpose.

%
%%%%%%%%%%%%%%%%%%%%%%%%%%%%%%%
%%%%% ACKNOWLEDGEMENTS %%%%%
%%%%%%%%%%%%%%%%%%%%%%%%%%%%%%%
%
\begin{acknowledgements}
We thank the anonymous reviewer for their comprehensive report, and
F.~M. Rica and M.~R. Zapatero Osorio for helpful comments.
We acknowledge financial support from the Agencia Estatal de Investigaci\'on of the Ministerio de Ciencia e Innovaci\'on and the ERDF ``A way of making Europe'' through projects
PID2022-138855NB-C31, % Gorgas'
PID2022-137241NB-C4[1:2], and % CAB+UCM-CARMENES
PID2020-112949GB-I00, % CAB SVO 
and from the European Commission Framework Programme Horizon 2020 Research and Innovation through the ESCAPE project under grant agreement no.~824064.
This research made use of the NASA's Astrophysics Data System Bibliographic Services and the Exoplanet Archive, which is operated by the California Institute of Technology, under contract with the National Aeronautics and Space Administration under the Exoplanet Exploration Program, 
the Washington Double Star catalogue maintained at the U.S. Naval Observatory,
the Simbad database \citep{wenger00}, 
the VizieR \citep{ochsenbein00} catalogue access tool, and the Aladin sky atlas \citep{bonnarel00} at the Centre de donn\'ees astronomiques de Strasbourg (France).
\end{acknowledgements}
%

%
%%%%%%%%%%%%%%%%%%%%%%%%%%%%%%%%%%%%%%%
%%%%%%%% Bibliography %%%%%%%%
%%%%%%%%%%%%%%%%%%%%%%%%%%%%%%%%%%%%%%%
%
%\begin{thebibliography}{}
\bibliographystyle{aa.bst}
\bibliography{mnemonic,biblio}
%\end{thebibliography}

\begin{appendix}

\section{Remarkable systems}
\label{sec:remarkable_systems}

\paragraph{HD~1466, two M dwarfs, and an early-L ultracool dwarf.}
Before our analysis, this system was made of the young late F-type star HD~1466, which hosts an astrometrically-inferred Jovian exoplanet \citep{mesa22}, in the Tucana-Horologium association and a wide M5.5\,V companion at about 30\,arcmin \citep{gaiacollaboration21b}.
We added to the system, which became quadruple, an early M dwarf and and L1 ultracool dwarf already reported as members of the Tucana-Horologium association \citep{kraus14,gagne15} at about twice the projected angular separation of the wide M5.5\,V companion.
The ultracool dwarf is the latest, least massive system component reported in this work.

\paragraph{HD~88072~AB and LP~609--39.}
The bright, close ($\rho$\,$\sim$\,3.7\,arcsec) pair HD~88072~AB was reported by \citet{gaiacollaboration21b}, while the planet ($\rho_{\rm planet}$\,$\sim$\,0.39\,arcsec calculated from the semi-major axis $a$ and distance $d$) was reported by \citet{feng22}.
To this compact, relatively new system we added another new component, namely LP~609--39, which is a poorly investigated $\sim$M5.5\,V companion at an extremely wide separation of about 80\,arcmin.

\paragraph{HD~134606 and L~72--1~[AB] (WDS~15154--7032).}
It was made of the three-exoplanet host star HD~134606 \citep{mayor11} and an $\sim$M3.5\,V companion at about 1\,arcmin, namely L~72--1.
Interestingly, the wide pair was discovered by an amateur astronomer \citep{rica12}.
As discussed in Sect.~\ref{close_binaries_without_Gaia}, L~72--1~[AB] displays a bimodal distribution of the \textit{Gaia} DR3 $G$-band light curve, which is indicative of a close binarity of the order of 0.2\,arcsec.
Furthermore, in his personal notes, F.\,M.~Rica also annotated a possible close binarity of L~72--1 based on the difference between its $V$-band absolute and apparent magnitudes.
Furthermore, the \textit{Gaia} DR3 $d$ value of L~72--1~[AB] by \citet{finch18} in Table~\ref{tab:systems1} is affected by close multiplicity.
%Therefore, the binary system became a hierarchical triple.

\paragraph{CD--24~12030 and two M-dwarf companions.}
The K2.5\,V-type primary star has received little attention since the tabulation by \citet{thome1892} in his Cordoba Durchmusterung (e.g. \citealt{degeus90,cruzalebes19,trifonov20}). 
It hosts a transiting planet discovered by K2 \citep{zink21,christiansen22}.  
The two co-moving companions at about 3.7\,arcmin, which are separated by 10.2\,arcsec between them (and 46.6\,deg of position angle), have not been previously discussed in the literature. 
The abnormal radial velocity of the faintest component is probably due to the orbital effect of the close binary.

\paragraph{HD~222259\,AB and 2MASS J23321028--6926537 (WDS~23397--6912).}
The system was originally made of the bright, close binary system HD~222259\,A and B, which are late G and early K dwarfs, respectively.
The pair, separated by about 5\,arcmin, was resolved for the first time by H.\,C.~Russell at the end of the 19th century, and is best known in the literaure as DS~Tuc~AB.
The primary, DS~Tuc~A, hosts an inflated transiting planet \citep{newton19,benatti19}.
We added a $\sim$M5\,V wide companion at about 43\,arcmin, namely 2MASS J23321028--6926537, which is also a member of Tucana-Horologium \citep{gagne15,ujjwal20}.
Actually, \citet{newton19} noticed the co-moving, co-distant candidate companion, but given its very wide separation they classified it as ``likely [...] unbound member of Tuc-Hor and not a bound companion of DS~Tuc''.
Its binding energy is $|U_g| \sim 4.6\,10^{33}$\,J, comparable to other wide binaries reported in the literature.

\paragraph{LP~141--14 and two M-dwarf companions.}
LP~141--14 is a white dwarf \citep[spectral type DC\,D,][]{dupuis94} located in a triple system, along with stars G 229--20A and G 229--20B, which are two M dwarfs with very similar individual masses and a combined mass of about 0.65\,M$_{\odot}$ \citep{stephan21}. 
The separation of LP~141--14 with the two close M dwarfs is 43.7\,arcsec (1082\,au), while the separation between the planet \citep[LP~141--14\,b,][]{vanderburg20} and the white dwarf it is orbiting around is 0.0204\,au. 
Together with the circumbinary planet RR\,Cae\,b, LP~141--14\,b is the only planet orbiting a white dwarf in our sample.

\section{Long tables}
\label{sec:sample_of_work}

\tiny

\captionof{table}{Exoplanet host stars at $d <$ 100\,pc.}
\tablefirsthead{\toprule \midrule 
  Star & $\alpha$ (J2000) & $\delta$ (J2000) & $d$  \\ 
    & (hh:mm:ss.ss) & (dd:mm:ss.s) & (pc) \\
  \midrule}
\tablehead{
\multicolumn{4}{c}
%{\tablename\ \thetable\: Non-duplicated exoplanet... \textit{(continued).}} \\
{\small{{\bf Table~\thetable :}} \normalsize{Exoplanet host stars at $d <$ 100\,pc \textit{(continued).}}} \\
\\
\toprule
\midrule
Star & $\alpha$ (J2000) & $\delta$ (J2000) & $d$  \\ 
  & (hh:mm:ss.ss) & (dd:mm:ss.s) & (pc) \\
\midrule}
\tabletail{
\midrule \multicolumn{4}{r} {\vspace{3pt}} \\  
}
\tablelasttail{
\bottomrule {}
}
%\xentrystretch{0.09}
\setlength{\extrarowheight}{3pt}
\begin{center}
\begin{xtabular}{@{\hspace{1mm}}p{45mm}@{\hspace{1mm}}c@{\hspace{1mm}}c@{\hspace{2mm}}c@{\hspace{4mm}}}
\label{tab:sample}

CD--55 9423 & 00:00:44.91 & --54:49:49.9 & 61.9 \\
HD 142 & 00:06:19.18 & --49:04:30.7 & 26.2 \\
G 158--27 & 00:06:43.20 & --07:32:17.0 & 4.8 \\
HD 457 & 00:08:59.68 & --39:44:13.8 & 55.3 \\
CD--27 16  & 00:09:04.36 & --27:07:19.7 & 23.8 \\
HD 564 & 00:09:52.82 & --50:16:04.2 & 51.2 \\
CD--39 31 & 00:11:41.65 & --39:13:47.8 & 34.9 \\
HD 984 & 00:14:10.25 & --07:11:56.8 & 45.7 \\
\end{xtabular}
\end{center}

\newpage
\onecolumn
\centering
\tiny

\begin{longtable}{lccccl}
 \caption{Multiple systems with one or more ultracool dwarfs and without proper planets.}\\
 \\
 \noalign{\smallskip}
 \hline \hline
 \noalign{\smallskip}
 Star & $\alpha$\,(J2000) & $\delta$\,(J2000) & \textit{G} & Spectral &  Reference of UCD discovery \\
 & (hh:mm:ss.ss) & (dd:mm:ss.s) & (mag) & type & \\
 \noalign{\smallskip}
 \hline
 \noalign{\smallskip} 
 \endfirsthead
 \caption{Multiple systems with one or more ultracool dwarfs and without proper planets \textit{(continued).}}\\
 \\
  \hline \hline
 \noalign{\smallskip}
 Star & $\alpha$\,(J2000) & $\delta$\,(J2000) & \textit{G} & Spectral &  Reference of  UCD discovery \\
 & (hh:mm:ss.ss) & (dd:mm:ss.s) & (mag) & type & \\
\noalign{\smallskip}
\endhead
\hline
\noalign{\smallskip}
\endfoot
 HD 984[A] & 00:14:10.25 & --07:11:56.8 & 7.2 & F7 & \\
 HD 984[B] & 00:14:10.25 & --07:11:56.8 & ... & M6.5 & \citet{johnsongroh17} \\
\noalign{\smallskip}
\hline
\noalign{\smallskip}
 DENIS J004135.3-562112[A] & 00:41:35.39 & --56:21:12.9  & 16.4 & M6.5 & \\
 DENIS J004135.3-562112[B] & 00:41:35.39 & --56:21:12.9  & ... & M9.0 & \citet{reiners10} \\
\noalign{\smallskip}
\hline
\noalign{\smallskip}
 HD 4747[A] & 00:49:26.77 & --23:12:44.9 & 7.0 & G8/K0 & \\
 HD 4747[B] & 00:49:26.77 & --23:12:44.9 & ... & T1 & \citet{crepp16} \\
\noalign{\smallskip}
\hline
\noalign{\smallskip}
 SCR J0103--5515[A] & 01:03:35.83 & --55:15:57.0 & 13.9 & M5.0 &  \\
 SCR J0103--5515[B] & 01:03:35.83 & --55:15:57.0 & ... & M5.0 &  \\
\end{longtable}

\newpage

\begin{table*}[h]
 \centering
 \caption[]{Astrometry of systems with new common $\mu$ and $\pi$ companions.}
  \medskip
  \medskip
 \scalebox{0.83}[0.83]{
 \begin{tabular}{lccccc}
 \hline \hline
 \noalign{\smallskip}
 Star & $\pi$ & $\mu_{\alpha}\cos\delta$ & $\mu_\delta$ & $\mu_{\rm ratio}$ & $\Delta$PA \\
  & (mas) & (mas\,a$^{-1}$) & (mas\,a$^{-1}$) &  & (deg) \\
 \noalign{\smallskip}
 \hline
 \noalign{\smallskip} 
* HD 1466 & 23.36 $\pm$ 0.02  &  90.05 $\pm$ 0.02  &  --59.21 $\pm$ 0.02  &    &    \\
UCAC3 52--533 & 20.97 $\pm$ 0.01  &  79.94 $\pm$ 0.02  &  --54.33 $\pm$ 0.02  &  0.10  &  0.88  \\
UCAC3 53--724 & 23.18 $\pm$ 0.08  &  91.47 $\pm$ 0.08  &  --54.92 $\pm$ 0.08  &  0.04  &  2.34  \\
2MASS J00191296--6226005 & 20.60 $\pm$ 1.67  &  88.10 $\pm$ 1.63  &  --54.50 $\pm$ 1.43  &  0.05  &  1.58  \\
 \noalign{\smallskip}
 \hline
 \noalign{\smallskip} 
* HD 27196 & 17.40 $\pm$ 0.01  &  --25.77 $\pm$ 0.01  &  --116.43 $\pm$ 0.02  &    &    \\
2MASS J04164536--2646205 & 17.40 $\pm$ 0.03  &  --24.52 $\pm$ 0.03  &  --118.49 $\pm$ 0.03  &  0.02  &  0.79  \\
\hline
 \end{tabular}
 } % \scalebox
 \label{tab:new_detected_systems_astrometry}  
\end{table*}

\newpage

\begin{longtable}{@{\hspace{1mm}}l@{\hspace{0.5mm}}l@{\hspace{0.5mm}}l@{\hspace{0.5mm}}c@{\hspace{0mm}}c@{\hspace{0.5mm}}c@{\hspace{0mm}}c@{\hspace{0.5mm}}c@{\hspace{0.5mm}}c@{\hspace{1mm}}l@{\hspace{0mm}}}
\caption{Basic data of the identified multiple stellar systems with planets.}\\
\\
\noalign{\smallskip}
\hline \hline
\noalign{\smallskip}
WDS & Disc. code / & Star$^{(b)}$ & $\alpha$ (J2000) & $\delta$ (J2000) & $d$ & $\rho^{\,(c)}$ & $\theta^{\,(c)}$ & \textit{s} & Refs.$^{(a)}$ \\
 & Disc. ref.$^{(a)}$ & & (hh:mm:ss.ss) & (dd:mm:ss.s) & (pc) & (arcsec) & (deg) & (au) & \\
\noalign{\smallskip}
\hline
\noalign{\smallskip}
\endfirsthead
\caption{Basic data of the identified multiple stellar systems with planets \textit{(continued).}}\\
\\
\hline \hline
\noalign{\smallskip} 
WDS & Disc. code / & Star$^{(b)}$ & $\alpha$ (J2000) & $\delta$ (J2000) & $d$ & $\rho^{\,(c)}$ & $\theta^{\,(c)}$ & \textit{s} & Refs.$^{(a)}$ \\
 & Disc. ref.$^{(a)}$ & & (hh:mm:ss.ss) & (dd:mm:ss.s) & (pc) & (arcsec) & (deg) & (au) & \\
\noalign{\smallskip}
\endhead
\hline
\noalign{\smallskip}
\endfoot 
00007--5450 &  & * CD--55 9423[A] & 00:00:44.91 & --54:49:49.9 & 61.87$\pm$0.04 &  &  &  &  \\
 & TOI 129 & CD--55 9423B & 00:00:44.65 & --54:49:46.8 & 61.34$\pm$0.26 & 3.813 & 322 & 235.9 & This work \\
 & KPP2838 & Gaia DR3 4923848227231122688 & 00:00:47.59 & --54:50:59.8 & 63.74$\pm$0.71 & 73.62 & 162 & 4554 & This work \\
\noalign{\smallskip}
\hline 
\noalign{\smallskip}
00063--4905 &  & * HD 142 & 00:06:19.18 & --49:04:30.7 & 26.18$\pm$0.02 &  &  & &   \\
 & HDO 180 & HD 1428 & 00:06:19.15 & --49:04:34.9 & 26.16$\pm$0.02 & 3.883 & 186 & 101.67 & This work \\
\noalign{\smallskip}
\hline 
\noalign{\smallskip}
00163--7951 &  & * HD 1237[A] & 00:16:12.68 & --79:51:04.2 & 17.55$\pm$0.01 &  &  &  & \\
 & GJ 3021 & HD 1237B & 00:16:12.57 & --79:51:00.4 & 17.59$\pm$0.03 & 3.985 & 352 & 69.96 & This work \\
\noalign{\smallskip}
\hline 
\noalign{\smallskip}
00184+4401 &  & * GX And & 00:18:22.89 & +44:01:22.6 & 3.56$\pm$0.01 &  &  &  &   \\
\end{longtable}
  \tablefoot{
    \tablefoottext{a}{WDS discoverer codes are written in uppercase, and literature references in lowercase. References: 
    Aka23: \citet{akana23};
    App23: \citet{apps23};
    Bai50: \citet{baize50}; 
    Bar20: \citet{barnes20}; 
    Ber23: \citet{bertini23}; 
    Bru98: \citet{bruch98}; 
    Car20: \citet{carleo20}; 
    Cha11: \citet{chauvin11}; 
    Chr22: \citet{christian22}; 
    Cla22: \citet{clark22}; 
    Dea14: \citet{deacon14};
    Des23: \citet{desidera23}; 
    Egg07: \citet{eggenberger07}; 
    Elb21: \citet{elbadry21}; 
    Fah12: \citet{faherty12}; 
    Far14; \citet{farrington14}; 
    Fei19: \citet{feinstein19}; 
    Fen22: \citet{feng22};
    Fon21: \citet{fontanive21};  
    Gai21: \citet{gaiacollaboration21b};
    Hei74: \citet{heintz74}; 
    Jus19: \citet{justesen19}; 
    Ker19: \citet{kervella19}; 
    Krz84: \citet{krzeminski84};
    Les21: \citet{lester21};
    Lod14: \citet{lodieu14};
    Ma16 : \citet{ma16}; 
    Mal12: \citet{malkov12}
    Mar20: \citet{marocco20}; 
    Mic21: \citet{michel21}; 
    Mug19: \citet{mugrauer19}; 
    Mug20: \citet{mugrauer20}; 
    Neu07: \citet{neuhauser07}; 
    Oh17: \citet{oh17}; 
    Pie23: \citet{pierens23}; 
    Pou04: \citet{pourbaix04};
    Prs11: \citet{prsa11}; 
    Ram09: \citet{ramm09};
    Rob15: \citet{roberts15}; 
    Rod16: \citet{rodigas16}; 
    Roe12: \citet{roell12}; 
    Ser22: \citet{serrano22};
    Su21:  \citet{su21}; 
    Udr02: \citet{udry02};
    Van19: \citet{vanderburg19}; 
    Wit16: \citet{wittrock16}; 
    Zho22: \citet{zhou22}; 
    Zie20: \citet{ziegler20};
    Zuc03: \citet{zucker03}.}
    \tablefoottext{b}{The exoplanet(s) host star is marked with an asterisk.}
    \tablefoottext{c}{We do not tabulate $\theta$ when $\rho<1$\,arcsec.
    Besides, we tabulate neither $\rho$ nor $\theta$ for the circumbinary planets Kepler~16 and RR~Cae.}
    }  
\end{appendix}
\end{document}